\documentclass[12pt,twoside]{mitthesis}
\usepackage{lgrind}

\usepackage{indentfirst}
\usepackage{cmap}
\usepackage{blindtext}
\usepackage[T1]{fontenc}
\usepackage{amsmath}
\usepackage{graphicx}
\usepackage{bm}
\usepackage{subfig}
\usepackage[usenames, dvipsnames]{color}
\graphicspath{{images/}}
\usepackage[backend=bibtex,style=ieee,natbib=true]{biblatex}
\begin{document}

% -*-latex-*-
% 
% For questions, comments, concerns or complaints:
% thesis@mit.edu
% 
%
% $Log: cover.tex,v $
% Revision 1.8  2008/05/13 15:02:15  jdreed
% Degree month is June, not May.  Added note about prevdegrees.
% Arthur Smith's title updated
%
% Revision 1.7  2001/02/08 18:53:16  boojum
% changed some \newpages to \cleardoublepages
%
% Revision 1.6  1999/10/21 14:49:31  boojum
% changed comment referring to documentstyle
%
% Revision 1.5  1999/10/21 14:39:04  boojum
% *** empty log message ***
%
% Revision 1.4  1997/04/18  17:54:10  othomas
% added page numbers on abstract and cover, and made 1 abstract
% page the default rather than 2.  (anne hunter tells me this
% is the new institute standard.)
%
% Revision 1.4  1997/04/18  17:54:10  othomas
% added page numbers on abstract and cover, and made 1 abstract
% page the default rather than 2.  (anne hunter tells me this
% is the new institute standard.)
%
% Revision 1.3  93/05/17  17:06:29  starflt
% Added acknowledgements section (suggested by tompalka)
% 
% Revision 1.2  92/04/22  13:13:13  epeisach
% Fixes for 1991 course 6 requirements
% Phrase "and to grant others the right to do so" has been added to 
% permission clause
% Second copy of abstract is not counted as separate pages so numbering works
% out
% 
% Revision 1.1  92/04/22  13:08:20  epeisach

% NOTE:
% These templates make an effort to conform to the MIT Thesis specifications,
% however the specifications can change.  We recommend that you verify the
% layout of your title page with your thesis advisor and/or the MIT 
% Libraries before printing your final copy.
\title{Knotted optical vortex lines in nonlinear saturable medium}

\author{Alfarabi Issakhanov}
% You can use the \\ command to list multiple previous degrees
%       \prevdegrees{B.S., University of California (1978) \\
%                    S.M., Massachusetts Institute of Technology (1981)}
\department{Department of Physics}

% If the thesis is for two degrees simultaneously, list them both
% separated by \and like this:
% \degree{Doctor of Philosophy \and Master of Science}
\degree{Master of Science in Physics}

% As of the 2007-08 academic year, valid degree months are September, 
% February, or June.  The default is June.
\degreemonth{June}
\degreeyear{2018}
\thesisdate{June 12, 2018}

%% By default, the thesis will be copyrighted to MIT.  If you need to copyright
%% the thesis to yourself, just specify the `vi' documentclass option.  If for
%% some reason you want to exactly specify the copyright notice text, you can
%% use the \copyrightnoticetext command.  
%\copyrightnoticetext{\copyright IBM, 1990.  Do not open till Xmas.}

% If there is more than one supervisor, use the \supervisor command
% once for each.
\supervisor{Anton S. Desyatnikov}{Full Professor}

\chairman{Vassilios D. Tourassis}{Dean, School of Science and Technology}

% Make the titlepage based on the above information.  If you need
% something special and can't use the standard form, you can specify
% the exact text of the titlepage yourself.  Put it in a titlepage
% environment and leave blank lines where you want vertical space.
% The spaces will be adjusted to fill the entire page.  The dotted
% lines for the signatures are made with the \signature command.
\maketitle

% The abstractpage environment sets up everything on the page except
% the text itself.  The title and other header material are put at the
% top of the page, and the supervisors are listed at the bottom.  A
% new page is begun both before and after.  Of course, an abstract may
% be more than one page itself.  If you need more control over the
% format of the page, you can use the abstract environment, which puts
% the word "Abstract" at the beginning and single spaces its text.

%% You can either \input (*not* \include) your abstract file, or you can put
%% the text of the abstract directly between the \begin{abstractpage} and
%% \end{abstractpage} commands.

% First copy: start a new page, and save the page number.
\cleardoublepage
% Uncomment the next line if you do NOT want a page number on your
% abstract and acknowledgments pages.
% \pagestyle{empty}
\setcounter{savepage}{\thepage}
\begin{abstractpage}
In last 50 years, a significant progress was noticed in medicine, communications and entertainment. Such advanced development of these fields was directly related to ability of controlling light. Photonics is exactly about this ability. At the present time, photonics is walking together with a fundamental physical concept, optical soliton. Optical solitons are shape-preserving laser beams. They are found potentially useful in data transmission, which is very significant nowadays. Hence, research in the field of optical solitons is still a vital issue. 
\par
% During the propagation of solitons in optical medium, they might be affected by random perturbations. 
When optical soliton is perturbed in a specific manner, there appear zeros of optical field around the soliton, which are called optical vortices. In general optical vortices are lines in space. Hence, we might expect them to become knotted. Knotting optical vortices around perturbed seems spontaneous and cannot be directly predicted. To explain this phenomenon, a similar system is constructed based on perturbation theory. In this system, however, we have a mathematical problem which yet lacks a full understanding. We address this problem by introducing concepts from three disciplines: laser physics, knot theory and singular optics. 
% \par
% Spatial optical solitons are shape-preserving continuous wave laser beams propagating in nonlinear medium. When optical soliton is perturbed, there appear optical vortices around the soliton, which are zeros of optical field.
% There is a basic understanding of this phenomenon. 
% \par
% A model, describing behaviour of laser beams during propagation, is a Nonlinear Schroedinger equation, which deterministic.
% Even though this system is described by deterministic equation, occurrence of optical vortices seems spontaneous and cannot be directly predicted. To avoid it, there is a way to generate similar system based on perturbation theory. However, in this case we have not well-defined mathematical problem. And this thesis is exactly about it. We address this problem by introducing concepts from three disciplines: laser physics, knot theory and nonlinear optics.
\par
We believe that understanding the mechanism underlying spontaneous knotting of optical vortices will be a step forward in other systems too, such as quantum turbulence in superfluids and formation of optical vortices around other types of optical solitons.
\end{abstractpage}

% Additional copy: start a new page, and reset the page number.  This way,
% the second copy of the abstract is not counted as separate pages.
% Uncomment the next 6 lines if you need two copies of the abstract
% page.
% \setcounter{page}{\thesavepage}
% \begin{abstractpage}
% \input{abstract}
% \end{abstractpage}

\cleardoublepage

\section*{Acknowledgments}
I would like to thank my advisor, Prof. Anton Desyatnikov, who guided me during these 2 years. I really enjoyed discussing with him, he explained everything with large enthusiasm and was always available to answer my questions, even when they were immature. When I lacked necessary enthusiasm needed for this work, he  shared new ideas and motivated me with great speeches. I must thank Volodymyr Biloshytskyi, for providing me with algorithms for simulations, and for sharing tips in numerical computations. I really enjoyed spending time with my good friend, Abay, especially when we were trying to find alternative and more intuitive explanations of fundamental physical concepts. And last, but not the least, I am grateful to my family, who supported my idea to study physics.

% This is the acknowledgements section.  You should replace this with your
% own acknowledgements. 

%%%%%%%%%%%%%%%%%%%%%%%%%%%%%%%%%%%%%%%%%%%%%%%%%%%%%%%%%%%%%%%%%%%%%%
% -*-latex-*-

% Some departments (e.g. 5) require an additional signature page.  See
% signature.tex for more information and uncomment the following line if
% applicable.
% \include{signature}
\pagestyle{plain}
% -*- Mode:TeX -*-
%% This file simply contains the commands that actually generate the table of
%% contents and lists of figures and tables.  You can omit any or all of
%% these files by simply taking out the appropriate command.  For more
%% information on these files, see appendix C.3.3 of the LaTeX manual. 
\tableofcontents
\newpage
\listoffigures
% \newpage
% \listoftables

%% This is an example first chapter.  You should put chapter/appendix that you
%% write into a separate file, and add a line \include{yourfilename} to
%% main.tex, where `yourfilename.tex' is the name of the chapter/appendix file.
%% You can process specific files by typing their names in at the 
%% \files=
%% prompt when you run the file main.tex through LaTeX.
\chapter{Introduction}
In this chapter, we briefly discuss the reasons which motivate us to work on presented problem. In later sections, we present theoretical background from laser physics, singular optics and knot theory, which are necessary to describe and analyze our problem.
\section{Motivation}
Development of communications and biophysics is directly related to progress in photonics. A problem, which we present in this work, is related to photonics and has a potential to contribute to fundamental sciences too, such as laser physics and singular optics. This problem was addressed in 2012 \cite{desyatnikov2012spontaneous}, and still is not fully analyzed. It combines two concepts, optical solitons and optical vortices. Both of these ingredients are important from theoretical and industrial points of view. 
\par
Optical vortices define a skeleton of optical fields, i.e. if optical vortex is embedded into optical field, then structure of the field changes accordingly, in order to be consistent with local structure around the vortex. Optical tweezer \cite{curtis2002dynamic,grier2003revolution} is an example of application of optical vortices, where they are used to control small objects, e.g. cells. 
\par
Optical soliton, which is a light with unchanging shape during propagation, is of big interest for two reasons: we do not see it in everyday life and they have a big potential to be used in industry. An idea, based on the specific type of optical solitons, was already commercialized in 2002 \cite{mitschke2012recent}. It was reported that a different type of optical solitons can behave as a waveguide \cite{shih1996circular}.
\par
Looking at applicability of optical vortices and optical solitons, we believe that a problem addressed in this work, which combines both of them, deserves working on it.
% \par
% Below, a thesis outline is presented.
% \par
% In chapter 1, we present some basics from laser physics, singular optics and knot theory, because these are necessary ingredients of the posed problem.
% \par
% In chapter 2, we present a model, describing propagation of continuous wave laser beams in nonlinear medium, and analyze stability of solitons in saturable medium.
% \par
% In chapter 3,  we consider topology of optical vortices around soliton, which resulted from particular type of perturbation.
% \par
% In the final chapter, we discuss about possible further steps to tackle the problem.
\section{Laser beams}
% In 1917, Albert Einstein proposed a theory which would result in invention of lasers. However, it took more than 40 years to invent them. 
In 1960, a first evidence of continuous wave laser was reported \cite{First}. A laser is considered as a source of coherent electromagnetic radiation. Therefore, it can be analyzed through Maxwell equations (Eq.(1.1)-(1.4)) 
\begin{equation}
\nabla \times \boldsymbol H(\boldsymbol r,t)=\boldsymbol J(\boldsymbol r,t)+\frac{\partial \boldsymbol D(\boldsymbol r,t)}{\partial t}
\end{equation}
\begin{equation}
\nabla \times \boldsymbol E(\boldsymbol r,t)=-\frac{\partial \boldsymbol B(\boldsymbol r,t)}{\partial t}
\end{equation}
\begin{equation}
\nabla \cdot \boldsymbol D(\boldsymbol r,t)=\rho (\boldsymbol r)
\end{equation}
\begin{equation}
\nabla \cdot \boldsymbol B(\boldsymbol r,t)=0
\end{equation}
where $\boldsymbol J$ is the electrical current density, $\boldsymbol H$ is the magnetic field, $\boldsymbol E$ is the electric field, $\rho$ is the electric charge density, $\boldsymbol D$ is the electric flux density and $\boldsymbol B$ is the magnetic flux density related to the corresponding fields as 
\begin{equation}
\boldsymbol D=\epsilon \boldsymbol E, \boldsymbol B=\mu \boldsymbol H
\end{equation}
$\mu$ is the permeability and $\epsilon$ is the permittivity. 
% Even though Maxwell's equations look neat, they carry a big amount of physical information. %We consider a free-space propagation of EM waves. 
To present general description of laser beams, we assume that laser beam propagates in a vacuum. Using a vector identity
\newline $\nabla \times \nabla \times \boldsymbol E=\nabla(\nabla \cdot \boldsymbol E)-\nabla ^2 \boldsymbol E$, separate equation for electric field can be derived
\begin{equation}
\nabla ^2 \boldsymbol E(\boldsymbol r,t)=\epsilon_0 \mu_0 \frac{\partial^2 \boldsymbol E(\boldsymbol r,t)}{\partial t}
\end{equation}
Eq.(1.6) is a wave equation, and each component of $\boldsymbol E$ satisfies it. Therefore, we can rewrite it in a scalar form for each component
\begin{equation}
\nabla ^2 E_{x,y,z}(\boldsymbol r,t)=\epsilon_0 \mu_0 \frac{\partial ^2E_{x,y,z}(\boldsymbol r,t)}{\partial t^2}
\end{equation}
{We assume that electric field wave is monochromatic
\begin{equation}
E_{x,y,z}(\boldsymbol r,t)=A_{x,y,z}(\boldsymbol r)e^{-i\omega t},
\end{equation}
where $A_{x,y,z}$ are complex amplitudes. For simplicity, we work with single component, which we call $A$. By substituting it into a wave equation, we get the time-independent equation for $A$
\begin{equation}
\nabla ^2 A(\boldsymbol r)=-k^2 A(\boldsymbol r),
\end{equation}
where $k=\frac{\omega}{c}$. Eq.(1.9) is scalar time-independent Helmholtz equation. Assume that laser beam is directed in $z$-direction and complex amplitude $A$ is separated as
\begin{equation}
A(x,y,z)=\psi(x,y,z)e^{ikz},
\end{equation}
where $\psi(x,y,z)$ is a slowly-varying envelope in comparison with $e^{ikz}$, since $k$ is large for visible light. By plugging it into time-independent Helmholtz equation, we get
\begin{equation}
\nabla ^2 \psi + 2ik\psi_{z}=0
\end{equation}
Lastly, we apply paraxial approximation
\begin{equation}
\left|2k\frac{\partial \psi}{\partial z}\right|\gg\left|\frac{\partial ^2 \psi}{\partial z^2}\right|
\end{equation}
Our final equation is
\begin{equation}
\frac{\partial ^2 \psi}{\partial x^2}+\frac{\partial ^2 \psi}{\partial y^2}+2ik\frac{\partial \psi}{\partial z}=0
\end{equation}}
Eq.(1.13) is a paraxial equation describing slowly-varying envelope of a monochromatic, highly-directed wave. The term $paraxial$ is used because all of the
light must travel nearly parallel to the optical axis, in our case $z$-axis, in order for the beam to have a sufficiently slow $z$ dependence. Laser beams typically obey this approximation very well.
\newline
A paraxial equation has several orthonormal sets of solutions, one of them is \textit{Hermite-Gauss modes} (HG). They are found using method of separation of variables in Cartesian coordinates, and the general form of these solutions is
\begin{equation}
\begin{split}
\psi_{mn}(x,y,z)= E_0\frac{w_0}{w(z)}H_m\left(\frac{\sqrt[]{2}x}{w(z)}\right)H_n\left(\frac{\sqrt[]{2}y}{w(z)}\right)\exp\left[-\frac{x^2+y^2}{w^2(z)}\right]\times \\
\times \exp\left[i\frac{k(x^2+y^2)}{2R(z)}\right]\exp\left[-i(m+n+1)\Phi(z)\right]    
\end{split}
\end{equation}
where $H_m$ and $H_n$ are Hermite polynomials.
The $m=n=0$ case is called \textit{a fundamental mode}. It is of the form 
\begin{equation}
\psi(x,y,z)=E_0\frac{w_0}{w(z)}\exp\left(-\frac{x^2+y^2}{w^2(z)}\right)\exp\left[i\frac{k(x^2+y^2)}{2R(z)}\right]\exp\left[-i \Phi(z)\right]
\end{equation}
\begin{figure}[h] %!t
\centering
\includegraphics[width=2.5in]{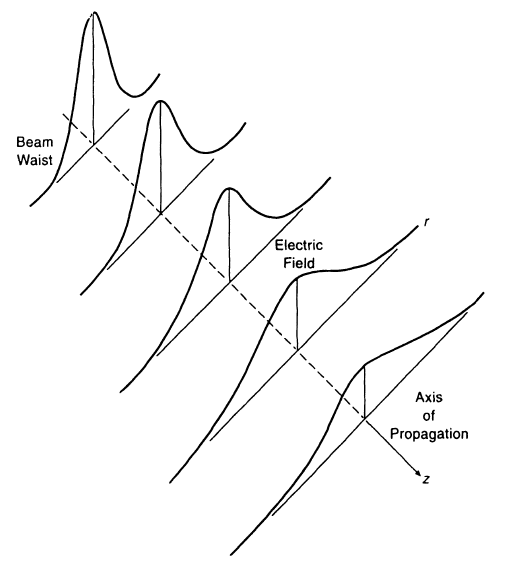}
\caption{Intensity distribution of fundamental mode at different $z$ \cite{goldsmith1998quasioptical}}
\label{bench}
\end{figure}
Quantities $w(z)$, $R(z)$ and $\Phi(z)$ have physical meanings. $R(z)$ is a radius of curvature of phase front, and the origin ($z=0$) is a position where the phase front has a radius of curvature $R=\infty$, i.e. behaves like a plane wave. $w(z)$ represents a size of beam spot, and at the origin, the spot size of the beam is $w_0$ and called \textit{a beam waist}, because it is the smallest spot size. 
A parameter $z_R$ is introduced, which is called \textit{a diffraction length}, in order to indicate how fast a beam diffracts.
When $z=z_R$, the size of a Gaussian beam has expanded by a factor of $\sqrt[]{2}$, and defined as $z_R=\frac{\pi w_0^2}{\lambda}$. For instance, consider two green light beams with waists $10 \mu\textrm{m}$ and $100\mu\textrm{m}$, then diffraction lengths are $1\textrm{mm}$ and $10\textrm{cm}$ respectively. Such dramatic difference is a result of quadratic dependence of $z_R$ on waist $w_0$. In other words, to get a small diffraction, we need a large waist.

% When $z\gg z_R$, $w(z)=w_0\frac{z}{z_R}$; since the relation between $w$ and $w_0$ becomes linear, a beam spreading angle $\theta$ can be defined as $\mathrm{tan}(\theta)=\frac{w}{z}=\frac{\lambda}{\pi w_0}$.
\begin{figure}[h] %!t
\centering
\includegraphics[width=4in]{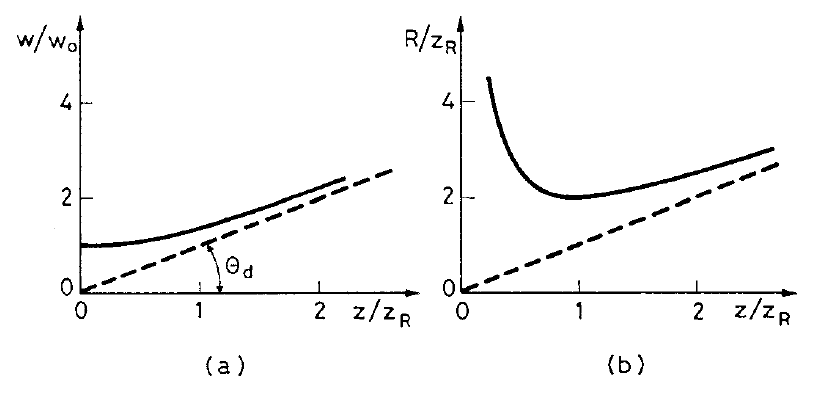}
\caption{ Normalized beam spot size $w$ (a) and radius of curvature $R$ (b) vs normalized $z$ \cite{svelto2009principles}}
\label{bench}
\end{figure}
Exact expressions for $R(z)$ and $w(z)$ are
\begin{equation}
w(z)=w_0\sqrt[]{1+z^2/z_R^2}
\end{equation}
\begin{equation}
R(z)=z+z_R^2/z
\end{equation}
The last parameter, $\Phi (z)=\mathrm{tan}^{-1}(\frac{z}{z_R})$ is called \textit{the geometrical Gouy phase}. It was shown to exist for any wave passing through a focus \cite{Feng:01}. 
% It can be proven that a field with minimal divergence and minimal transversal extension is a fundamental Gaussian beam \cite{pampaloni}.
\par
Applying separation of variables to paraxial equation in cylindrical coordinates, a different set of orthonormal solutions can be derived, which are called \textit{Laguerre-Gauss modes} (LG)
\begin{equation}
\begin{split}
\psi_{lp}(r,\theta,z)= E_0\frac{w_0}{w(z)}\left(\frac{\sqrt[]{2}r}{w(z)}\right)^{|l|} L_p^{|l|}\left(\frac{\sqrt[]{2}r}{w(z)}\right)\exp\left(-\frac{r^2}{w^2(z)}\right)\times \\ \times \exp\left(i\frac{kr^2}{2R(z)}\right)\exp(il\theta)\exp\left[-i(2|l|+p+1)\Phi(z)\right],
\end{split}
\end{equation}
where $p$ (radial index) is a positive integer, $l$ (azimuthal index) is an integer and $L_p^l$ is a generalized Laguerre polynomial. 
\begin{figure}[h] %!t
\centering
\subfloat[]{{\includegraphics[width=2.5in]{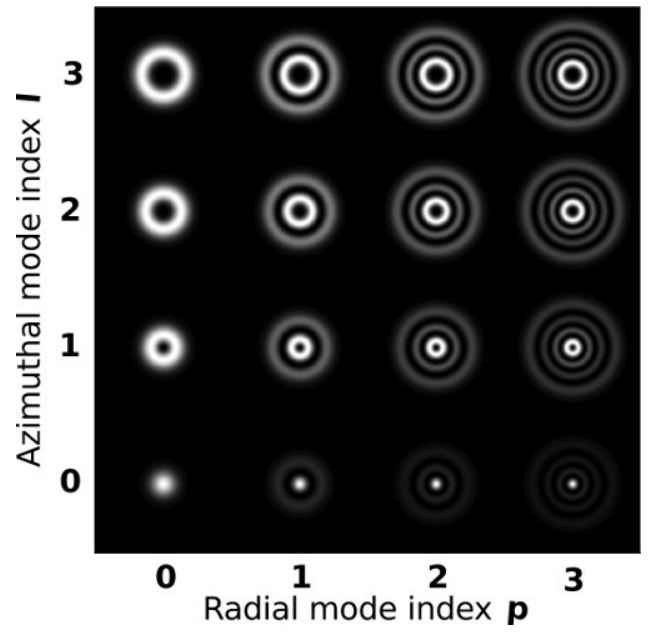} }}%
\qquad
\subfloat[]{{\includegraphics[width=2.5in]{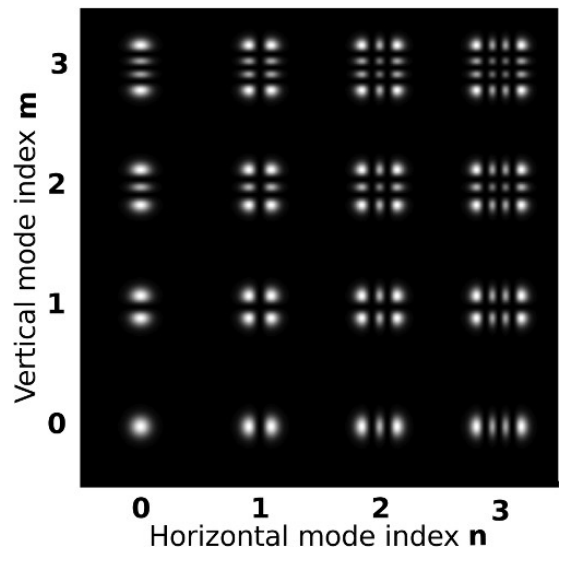} }}%
\caption{Intensity distributions of LG modes (a) and HG modes (b) up to 9th order \cite{carbone2013generation}}
\label{bench}
\end{figure}
The number $l$ defines number of full cycles of 2$\pi$ the phase is changing when one goes around the axis of propagation. 
% In addition, differently from Hermite-Gaussian beams, Laguerre-Gaussian modes have rotational symmetry
% along their propagation axis and carry an orbital angular momentum \cite{Intr_extr}.
Figure 1-3 presents intensity distributions for LG and HG modes up to 9th order. For $l=p=0$, a fundamental Gaussian mode is reconstructed. A more striking fact is that for $l\neq 0$, intensity vanishes on the propagation axis. Since both types of modes are separate sets of basis solutions, LG modes can be represented as a superposition of HG modes, as in Figure 1-4.
\begin{figure}[h] %!t
\centering
\includegraphics[width=3.5in]{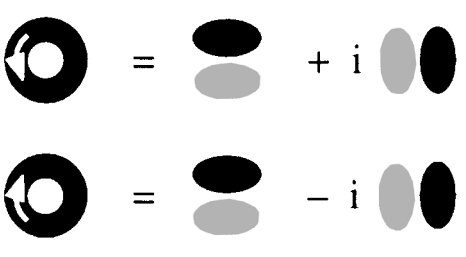}
\caption{HG$_{01}$ and HG$_{10}$ are superimposed to form LG$_{-1,0}$ and LG$_{1,0}$ modes \cite{Padgett:99}}
\label{bench}
\end{figure}
In case of LG modes, vanishing intensity is a consequence of interference.
However, a point where intensity vanishes is not a property related to LG modes only, it is even more generic. It can appear in a superposition of any waves at points of complete destructive interference.

\section{Optical vortices}
In the previous section, we described properties of laser beams. A laser emits light which can be described as a superposition of modes. Because of complete destructive interference, there may appear points in the beam where intensity vanishes. This phenomenon is generic in wave physics, and will be discussed here.
\par
We are interested in optical fields which can be expressed as a scalar field 
\begin{equation}
f(\boldsymbol r,t)=A(\boldsymbol r,t)\exp [iX(\boldsymbol r,t)],
\end{equation}
where $A(\boldsymbol r,t)$ and $X(\boldsymbol r,t)$ are real amplitude and phase respectively. At point of vanishing field, the phase $X$ is undefined; therefore, such points are called \textit{phase singularities}, or \textit{dislocations}. Second term is also used in crystallography, and this is not a coincidence. 
Consider LG mode $\psi_{1,0}$ with optical $z$-axis.
%logical error. you consider specific mode initially, then you analyze by assumptions.
There is a little change of field right in the neighborhood of phase singularity, therefore, change of amplitude in $z$ is negligible, i.e. $\frac{\partial \psi}{\partial z}\approx 0$ \cite{gbur2016singular}. Hence, resultant paraxial equation in the neighborhood of optical axis is
\begin{equation}
\frac{\partial^2 \psi}{\partial x^2}+\frac{\partial^2 \psi}{\partial y^2}=0,
\end{equation}
which is Laplace's equation. Its general solution expanded in Taylor series with only lowest order considered is
\begin{equation}
\psi=a(x+iy)+b(x-iy)
\end{equation}
For simplicity, if we consider $b=0$, then general field is $U(x,y,z)=a(x+iy)e^{ikz}$. If we represent a surface of constant phase, say 0, it is described by equation $\phi+kz=0$, where $\phi=\textrm{tan}^{-1}(y/x)$. Figure 1-5 illustrates this phase front, an it reminds a screw-type dislocation in crystal lattice.
\begin{figure}[h] %!t
\centering
\includegraphics[width=3.5in]{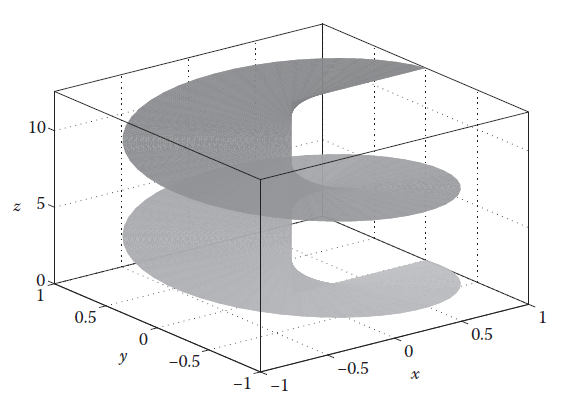}
\caption{A surface of constant phase 0 \cite{gbur2016singular}}
\label{bench}
\end{figure}
\par
Phase singularities with helical phase fronts around them are called \textit{optical vortices}. Optical vortices in LG modes, such as LG$_{-1,0}$ and LG$_{1,0}$ have to be distinguished. To do that, \textit{a winding number}, or \textit{topological charge} $m$ of  optical vortex is introduced as
\begin{equation}
m=\frac{1}{2\pi}\oint\nabla Xd\boldsymbol r,
\end{equation}
where $X$ is a phase, and integral is computed around an optical vortex along a closed non-intersecting path. Since $X$ changes by a multiple of $2\pi$ along this path, $m$ is an integer. In addition, this integral is computed in a conventional counter-clockwise fashion. Using Eq.(1.22), LG$_{-1,0}$ and LG$_{1,0}$ have topological charges -1 and 1 respectively.
\par
To detect optical vortices experimentally, intensity map is not sufficient because a region with extremely low intensity has a non-zero size and it is hard to measure location of optical vortex precisely. In addition, optical vortices of charge $\pm m$ might look identical. Thus, some special methods are necessary to detect them. An idea for one of the methods comes from the process of generation of optical vortices. In this method, by taking a plane wave along the $x$-axis, and interfering with Laguerre-Gauss beam of charge $m$, a resultant intensity distribution on a detector is 
\begin{equation}
I(x,\phi)=|\exp{(im\phi)}+\exp{(ik_{x}x)}|^2
\end{equation}
Figure 1-6 is an example of interference of different vortex beams with a plane wave.  
\begin{figure}[h] %!t
\centering
\includegraphics[width=4in]{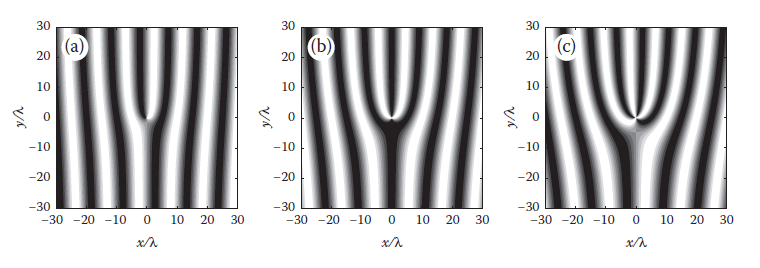}
\caption{Interference patterns of a plane wave and vortex beams with (a) $m=1$, (b) $m=2$, (c) $m=3$ \cite{gbur2016singular}}
\label{bench}
\end{figure}
We can observe that there are forks of interference fringes, which indicate the topological charge of optical vortex in the original vortex beam. 
\par
Another way to represent the envelope of optical field is
\begin{equation}
f(\boldsymbol r)=a(\boldsymbol r)+ib(\boldsymbol r)
\end{equation}
At point with vanishing field, real and imaginary parts vanish simultaneously. In other words, vortex is an intersection of two surfaces, which is a line.
Figure 1-7 is a phase map of a function, with pair of optical vortices in a plane.
\begin{figure}[h] %!t
\centering
\includegraphics[width=1.8in]{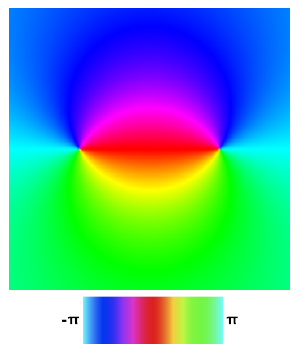}
\caption{Planar phase map of a function with two different phase singularities \cite{kingrp}}
\label{bench}
\end{figure}
This figure is a single cut, but it might be the case that if we combine many such planes, different vortex structures will occur, such as a loop, two infinite vortex lines, or it can even be knotted. As we consider in chapter 3, when medium is nonlinear, spontaneous knotting of optical vortices around the beam occurs. Therefore, it is reasonable to investigate topology of vortex lines, starting from the basics of knot theory.

\section{Knot theory}
Knot theory is a branch of topology, which started to develop two centuries ago.
Apart from the abstract topology of knot theory, only the specific case of three-dimensional knots will be considered in this section.
\par
A knot is a result of deformation of easily deformable string with no thickness, and with glued ends. In Figure 1-8(a), if a ring is deformed in such way that it does not intersect itself during deformation, then they are equivalent. Figure 1-8(b) illustrates a trefoil knot, which cannot be constructed from a ring only by continuous deformations. How do we know that they are not the same? This is a main aim of a knot theory, to find a way to categorize knots according to their topological properties.
\begin{figure}[h] %!t
\centering
\includegraphics[width=3in]{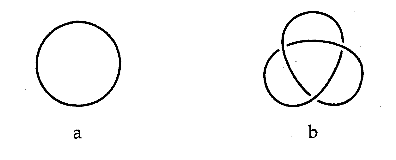}
\caption{Projections of (a) ring and (b) trefoil knot \cite{knotbook}}
\label{bench}
\end{figure}
\par
Figure 1-8 represents projections of a ring and trefoil knot. Every knot does not have a unique representation. For instance, we can take a ring with both hands and turn one hand by $\pi$. Although it is the same ring, its representation will be different from Figure 1-8(a). To prove that two different representations of the same knot are equivalent, \textit{Reidemeister moves} can be used. Essential idea of these moves is that we can transform representation of a certain knot without changing its topological properties. In \cite{reidemeister1927elementare}, it was proven that if there are two representations of a knot, then by using a series of these moves we can transform one projection to the other.
\par
There are many types of knots. However, the one which we will need later is \textit{a torus knot}, which lies  on a torus without crossing. Previously mentioned trefoil knot is a torus knot.
\begin{figure}[h] %!t
\centering
\subfloat[]{{\includegraphics[width=2in]{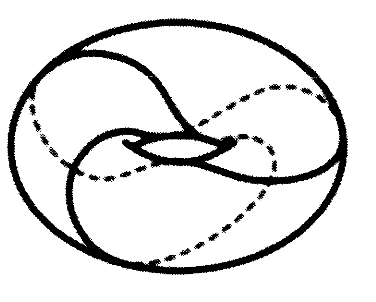} }}%
\qquad
\subfloat[]{{\includegraphics[width=2.2in]{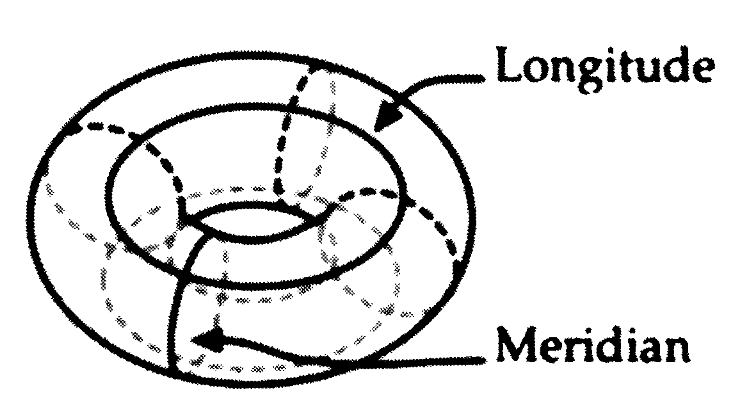} }}%
\caption{(a) Trefoil knot lying on a torus; (b) Meridian and longitude lines on a torus \cite{knotbook}}
\label{bench}
\end{figure}
\par
A curve that runs once the short way around the torus is \textit{a meridian curve}. A curve that runs once around the long way is called
\textit{a longitude curve}. 
The trefoil knot in Figure 1-9(a) wraps three
times meridianally around the torus and twice longitudinally. A trefoil knot is called (3, 2) torus knot, since it crosses longitude three times and meridian twice. Every torus knot is a ($p$, $q$)-torus knot for some pair of integers. In fact, the two integers will always be relatively prime \cite{knotbook}.
\par
Just by looking at projections of two different knots, it is difficult to tell them apart. To do that, a specific polynomial can be assigned to each projection \cite{knotbook,knottheory}. An advantage of these polynomials is that if two knots have different polynomials, then these knots are not equivalent. However, if two knots have the same polynomial, we cannot conclude anything about equivalence. There are several examples of different knots, having the same polynomials.
The first polynomial
associated to knots and links was due to J. Alexander in about 1928 \cite{alexander}. Mathematicians used the Alexander polynomial to distinguish
knots and links for the next 58 years. In particular, we are interested in torus knots, and fortunately, there is a general Alexander polynomial \cite{knottheory} for a $(p,q)$ torus knot
\begin{equation}
\Delta=\frac{(t^{pq}-1)(t-1)}{(t^p-1)(t^q-1)}
\end{equation}
$\Delta$ is a polynomial of $t$, which can be constructed by algorithm described in Alexander's original paper \cite{alexander}. Essential idea of construction: 
\begin{itemize}
\item let projection of knot have $n$ crossings, then there are $n+2$ regions (including outer)
\item around every crossing, there are 4 regions {$r_j$, $r_k$, $r_l$, $r_m$} (anti-clockwise)
\item construct sum of 4 regions, using coefficients $t,-t,+1,-1$ and equate it to $0$, e.g. $tr_j-tr_k+r_l-r_m=0$
\item $n$ equations with $n+2$ unknowns are formed. Construct corresponding matrix of dimensions $n\times (n+2)$. Delete any two columns and calculate the determinant of $n\times n$ matrix. Determinant is the Alexander polynomial
\end{itemize}
This polynomial does not take any input value, it only represents a specific knot. For example, Alexander polynomial of a trefoil knot is $\Delta(Trefoil)=1-t+t^2$.
\par
Coming back to optical vortices, which are zeros of optical field, one demonstrated way of constructing a function with a nodal set in the form of a knot was presented in \cite{kingrp}. Initial step is to construct braids, as a nodal set of some polynomial. Braid is a structure consisting of 2 or more strands. Suppose, braid is inserted inside the cylinder. A candidate describing a field with a nodal set in form of a braid of $N$ strands is polynomial
\begin{figure}[h] %!t
\centering
\subfloat[]{{\includegraphics[width=1in]{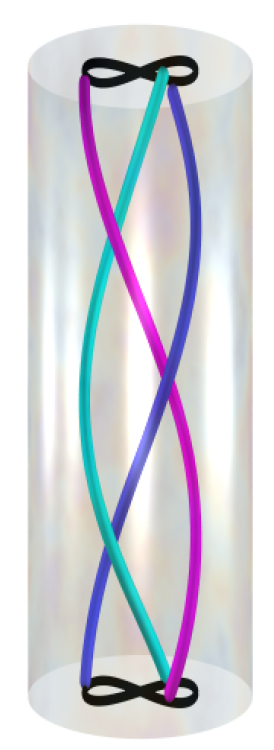} }}%
\qquad
\subfloat[]{{\includegraphics[width=2.2in]{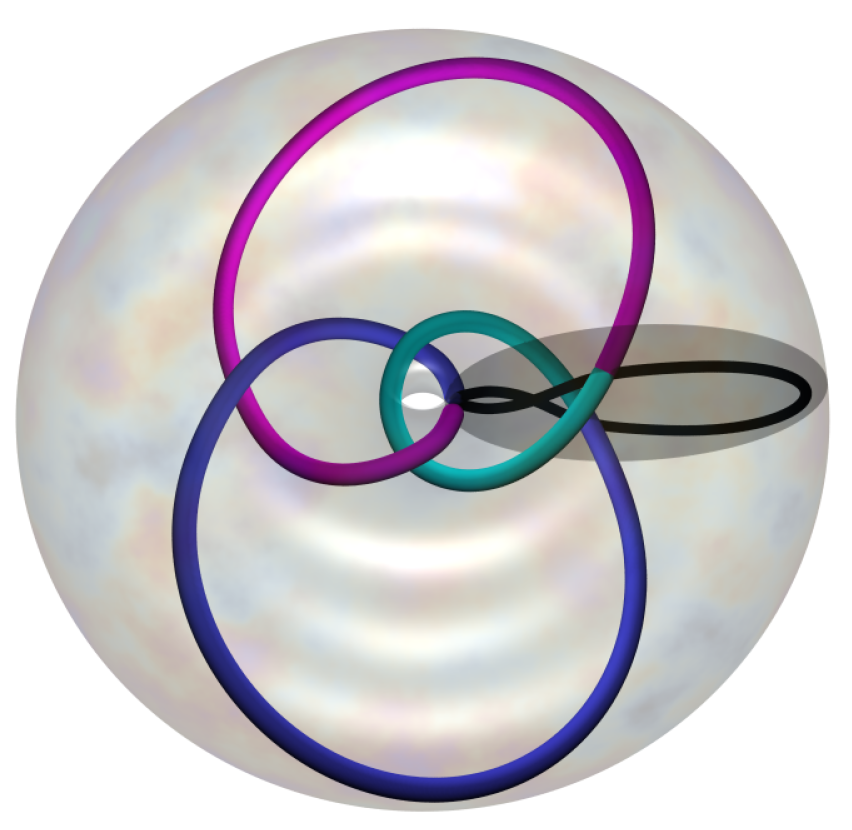} }}%
\caption{(a) Nodal set of Eq.(1.26) for $N=3$; (b) Knotted nodal set resulting from specific mapping $u(R,z)$, $v(R,z,\phi)$ \cite{kingrp}}
\label{bench}
\end{figure}
\begin{equation}
p_h(u)=\prod _{j=0}^{N-1}[u-s_j(h)],
\end{equation}
where $h$ is a height parameter and $s_j$ are periodic functions. For any $h$, there are $N$ roots in $u$ plane. For simplicity, we will consider only trigonometric $s_j$ functions. For instance, by taking $s_j(h)=\textrm{cos}(h-2\pi j/3)+i\textrm{sin}(2h-4\pi j/3)$ for $N=3$, Figure 1-10(a) represents braided nodal set. Out of this nodal set it is possible to construct a function whose nodal set is a knot. By setting $v=e^{ih}$, Figure 1-10(a) can be described by $q(u,v)=64u^3-12u(2v^2+2v^{*2}+3)-(14v^2+14v^{*2}+v^4-v^{*4})$. Then, using a substitution $u=\frac{R^2+z^2-1+2iz}{R^2+z^2+1}$ and $v=\frac{2Re^{i\phi}}{R^2+z^2+1}$, we connect the braids by corresponding ends, and get a real function whose nodal set is a figure-8 knot, as in Figure 1-10(b). We note that constructed nodal sets are not necessarily torus knots, e.g. figure-8 knot, which lies inside the torus (not on the surface). Based on above constructing algorithm, a knotted vortex was embedded in laser beams, which we will present in more details in chapter 3. 
\par
In this chapter, we presented some background from laser physics, singular optics and knot theory. In particular, we described an equation for continuous wave laser beam in vacuum, and sets of orthogonal solutions in different coordinates. Then, we presented phase singularities, which are zeros of optical field. Alexander polynomial was introduced, which is an invariant of a knot. Presented knowledge will be necessary to describe the problem of our interest.

%% This is an example first chapter.  You should put chapter/appendix that you
%% write into a separate file, and add a line \include{yourfilename} to
%% main.tex, where `yourfilename.tex' is the name of the chapter/appendix file.
%% You can process specific files by typing their names in at the 
%% \files=
%% prompt when you run the file main.tex through LaTeX.
\chapter{Optical solitons}
In this chapter, we introduce the nonlinear Schroedinger equation, which describes a propagation of laser beams in nonlinear medium. In particular, we are interested in Kerr  and saturable materials. Also, we will consider stationary solutions of nonlinear Schroedinger equation and their stability properties.
\section{Nonlinear Schroedinger equation}
In previous chapter, we derived a paraxial equation describing evolution of a slowly varying field envelope in a vacuum. 
% We can note that, it is a Schroedinger type equation.
If we consider a propagation of laser beams in a dielectric medium, it will induce
% an additional electric field, which is called 
a polarization field. In this case, an electric induction field $\boldsymbol D$ appearing in Maxwell's equations is defined as 
\begin{equation}
\boldsymbol D=\epsilon_0 \boldsymbol{\mathcal{E}} +\boldsymbol P
\end{equation}
For simplicity, we assume that light and induced polarization field are linearly polarized, then we have simplified expressions for $\boldsymbol{\mathcal{E}}=(\mathcal{E},0,0)$, $\boldsymbol P=(P,0,0)$ and hence $\boldsymbol D=(D,0,0)$. The same as before, we consider monochromatic electric field $\mathcal{E}(x,y,z,t)=e^{-i\omega t}E(x,y,z)+c.c.$ 
We expand polarization field $P$ in Taylor series
\begin{equation}
P=\epsilon_0\left[\chi^{(1)}\mathcal{E}+\chi^{(2)}(\omega)\mathcal{E}^2+\chi^{(3)}(\omega)\mathcal{E}^3+...\right],
\end{equation}
where $\chi^{(i)}$ is the $i$th-order optical susceptibility. At low intensities, higher order terms are negligible, and system is at the linear regime, i.e.
\begin{equation}
P=P_{lin}=\epsilon_0 \chi^{(1)}(\omega)\mathcal{E},
\end{equation}
Hence, the electric induction field is
\begin{equation}
D=\epsilon_0\mathcal{E}+P_{lin}=\epsilon_0n_0^2(\omega)\mathcal{E},
\end{equation}
where $n_0$ is the refractive index of the medium, defined as $n_0^2(\omega)=1+\chi^{(1)}(\omega)$. The only difference between Maxwell's equations in vacuum and in linear medium is the term $n_0^2$.
By the same procedure as in chapter 1, we get a scalar Helmholtz equation
\begin{equation}
\Delta E(x,y,z)+k^2_{0}E=0, \quad k_0^2=\frac{\omega^2}{c^2}n_0^2
\end{equation} 
When we consider more powerful beams, nonlinearities come into the play. The polarization field is 
\begin{equation}
P=P_{lin}+P_{nl},
\end{equation}
where $P_{nl}$ is contribution from remaining terms
\begin{equation}
P_{nl}=\epsilon_0\left[\chi^{(2)}(\omega)\mathcal{E}^2+\chi^{(3)}(\omega)\mathcal{E}^3+...\right]
\end{equation}
For the case of an isotropic medium, in which we are interested here, it can be proven that $\chi^{(2j)}=0$. Hence, for an isotropic medium in weakly-nonlinear regime, $P_{nl}\approx \chi^{(3)}\mathcal{E}^3$. An explicit expression for $P_{nl}$ is 
\begin{equation}
P_{nl}\approx\chi^{(3)}\left(3|E|^2Ee^{-i\omega t}+E^3e^{-3i\omega t}+c.c.\right)
\end{equation}
A second term has a frequency $3\omega$, and this phenomenon is known as \textit{a third-harmonic generation}. In general, effect from third-harmonic can be neglected \cite{fibich2015nonlinear}, therefore
\begin{equation}
P_{nl}\approx 3\chi^{(3)}|E|^2Ee^{-i\omega t}+c.c.=3\chi^{(3)}|E|^2\mathcal{E} 
\end{equation}
In this case, we have a modified electric induction field $D$
\begin{equation}
D=\epsilon_0\mathcal{E}+P_{lin}+P_{nl}=\epsilon_0n^2\mathcal{E},
\end{equation}
where $n$ is defined as
\begin{equation}
n^2=n^2\left(\omega,|E|^2\right)=n_0^2+\frac{3\chi^{(3)}|E|^2}{\epsilon_0}
\end{equation}
By introducing \textit{a Kerr coefficient} $n_2=\frac{3\chi^{(3)}}{4\epsilon_0n_0}$, the index of refraction can be rewritten as $n^2=n_0^2\left(1+\frac{4n_2}{n_0}|E|^2\right)$. Materials with such index of refraction are called \textit{Kerr materials}.
For many materials, $n_2$ is very small \cite{fibich2015nonlinear}. For instance,  water has $n_0\approx 1.33$ and $n_2\approx 10^{-22}m^2/V^2$. Hence, a nonlinear contribution induced by natural light sources (e.g. sunlight's $|E|\approx 10^3 \frac{V}{m}$) is $n_2|E|^2\approx10^{-16}$, i.e. $n_2|E|^2$ is negligible compared to $n_0$. From this example, we see that a powerful beam is the key to observe Kerr effect. 
Similarly, as for a linear dielectric, we can get an equation describing the propagation of linearly-polarized cw laser beam in Kerr material, known as \textit{scalar nonlinear Helmholtz equation} (NLH)
\begin{equation}
\Delta E(x,y,z)+k^2E=0, \quad k^2=k_0^2\left(1+\frac{4n_2}{n_0}|E|^2\right)
\end{equation}
By substituting $E=\psi e^{ik_0z}$ in the NLH and applying paraxial approximation, we get \textit{nonlinear Schroedinger equation} (NLS)
\begin{equation}
2ik_0\frac{\partial \psi}{\partial z}+\Delta_{\perp}\psi+k_0^2\frac{4n_2}{n_0}|\psi|^2\psi=0
\end{equation}
NLS shows that propagation of the beam is dependent on the effects of diffraction (2nd term) and Kerr nonlinearity (3rd term). As our next step in investigation of NLS, we will make it dimensionless, using appropriate transformations
\begin{equation}
x'=\frac{x}{r_0}, \quad y'=\frac{y}{r_0}, \quad z'=\frac{z}{2k_0r_0^2}, \quad \psi'=\frac{\psi}{E_c}
\end{equation}
where $r_0$ is the radius of input beam, $k_0r_0^2$ is the Rayleigh length and $|E_c|$ is some characteristic value. In dimensionless form of NLS below, primes are removed for simplicity 
\begin{equation}
i\frac{\partial \psi}{\partial z}+\Delta_{\perp}\psi+\alpha|\psi|^2\psi=0,
\end{equation}
where $\alpha=r_0^2k_0^2\frac{4n_2E_c^2}{n_0}$. When $\alpha\ll 1$ diffraction dominates over nonlinearity and propagation is weakly nonlinear; when $\alpha=O(1)$ diffraction and nonlinearity are of comparable magnitudes and the propagation is nonlinear and when $\alpha \gg 1$, Kerr effect dominates over diffraction. Hence, $\alpha$ is called \textit{a nonlinearity parameter}. Let $\psi$ be a solution of NLS, then there are several invariants
\begin{itemize}
\item power $P(z)=\int{|\psi|^2dxdy}$
\item Hamiltonian $H(z)=\int{|\nabla \psi|^2dxdy}-\frac{\alpha}{2}\int{|\psi|^4}dxdy$
\item linear momentum $M(z)=Im\int{(\psi\nabla\psi^*-\psi^*\nabla\psi)dxdy}$
\item angular momentum $L(z)=Im{\int(x,y)\times(\psi^*\nabla\psi-\psi\nabla\psi^*)dxdy}$
\end{itemize}
are conserved. These and other invariants are necessary in numerical simulations and investigation of solutions of NLS. NLS describes a propagation of continuous wave (cw) laser beam in a Kerr material, but the only difference between paraxial equation (linear Schroedinger) and NLS is the nonlinearity term. A positive Kerr nonlinearity ($n_2>0$)  corresponds to focusing of the beam propagating in such a medium \cite{fibich2015nonlinear}. 
% \textcolor{red}{To gain an intuitive explanation of this effect, geometrical optics is sufficient. Consider a Gaussian input beam
% \begin{equation}
% E_0(x,y)=E_ce^{-\frac{r^2}{r_0^2}}e^{i(kz-wt)}+c.c.
% \end{equation}
% entering Kerr material located in right half-space. Under the geometrical optics approximations, 
% \begin{equation}
% |E(x,y,z)|\approx E_ce^{-\frac{r^2}{r_0^2}}
% \end{equation}
% Electric field induces a change in the index of refraction stated before
% \begin{equation}
% n^2=n_0^2+4n_2n_0|E|^2\approx n_0^2+4n_2n_0E_c^2e^{-\frac{2r^2}{r_0^2}}
% \end{equation}
% We see that $n^2$ has a maximum value at $r=0$ and decreases with increasing $r$. Therefore, the beam rays tend to bend to the region of high index of refraction, i.e. to the center.} 
Since this focusing effect is caused by the input beam itself, such phenomenon is called \textit{self-focusing}. In the case of self-focusing effect being much stronger than diffraction, a catastrophic collapse might occur.
\begin{figure}[h] %!t
\centering
\includegraphics[width=1.8in]{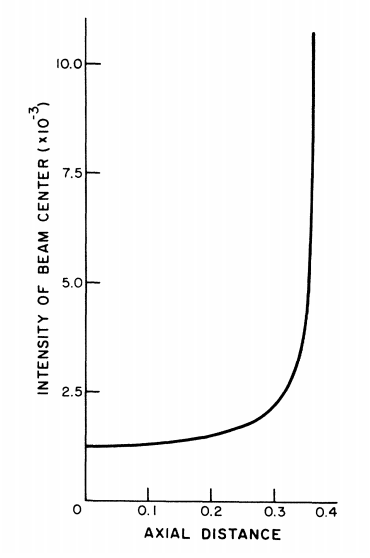}
\caption{Beam center intensity vs propagation distance \cite{kelley1965self}}
\label{bench}
\end{figure}
In 1965, a work by P.L.Kelley \cite{kelley1965self} suggested that diffraction during 2D propagation would not prevent from collapsing to a point. In Figure 2-1, taken from Kelley's paper, we see a dependence of intensity of the beam center on the propagation distance. In particular we see that intensity goes to infinity at a finite distance, which implies a catastrophic collapse. The reason for that is Kerr nonlinearity, it has no upper bound, which can result in damage of the medium. To prevent it, a different material can be used possessing a saturable nonlinearity. This type of nonlinearity has a big difference from a Kerr nonlinearity: as a function of intensity $I=|\psi|^2$, it has an upper bound, i.e. extreme self-focusing will not occur. Some materials, which will be introduced in the next section, can be modeled to have a saturable nonlinearity of the form $\frac{I}{1+I/I_s}$. This behavior is shown in Figure 2-2,
\begin{figure}[h] %!t
\centering
\includegraphics[width=2in]{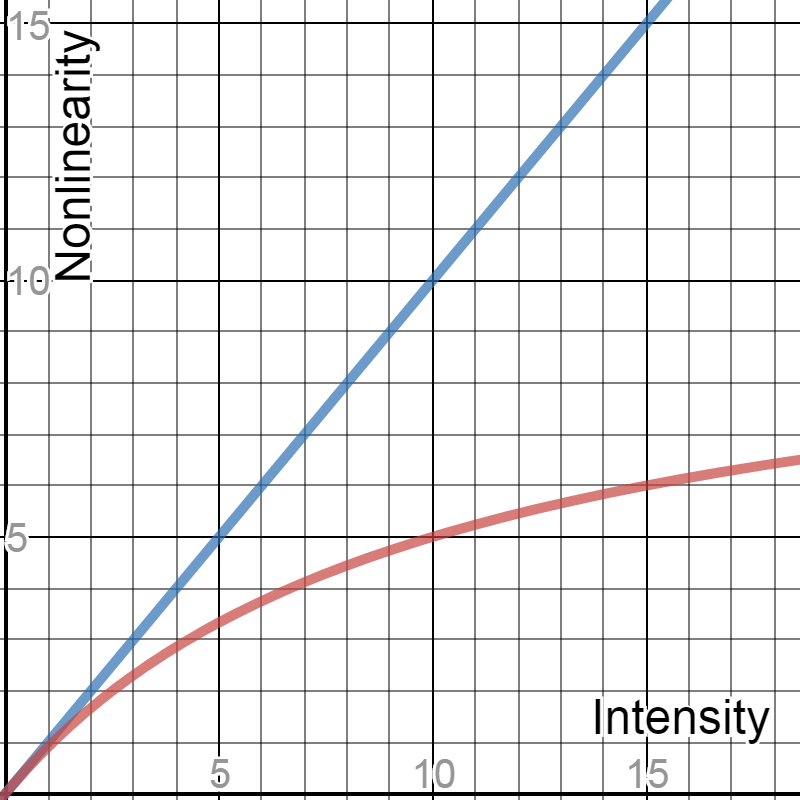}
\caption{Kerr (blue) and saturated (red) nonlinearities vs intensity}
\label{bench}
\end{figure}
the red line is a function of the form $\frac{x}{1+x/10}$ and the blue one is $x$. It is seen from the graph that saturable nonlinearity never exceeds Kerr nonlinearity and as intensity grows, a saturable nonlinearity approaches $I_s$, i.e. it saturates. 
% This type of nonlinearity provided new opportunities in optics, since then a light could be analyzed in a more controllable way.
\par
Above we considered the case, when Kerr effect dominates diffraction. The case, when nonlinearity and diffraction are comparable, or even compensate each other, is considered in the next section.
\section{Optical solitons in a medium with a saturable nonlinearity}
% In the previous section, we came to the conclusion that materials possessing Kerr nonlinearity will result in catastrophic collapse of an input beam. However we did not mention that
A wave tends to spread as it propagates, however, there is a way to generate non-diffracting waves, so that it would not even change its shape during propagation. Such waves are called \textit{solitons}. Soliton theory is a broad branch of physics, here we are interested in spatial optical solitons, i.e. non-spreading cw laser beams. From the previous section, we know that this type of waves is a result of a combination of two effects: diffraction and induced nonlinearity. Nonlinearity is not the only way to compensate diffraction. In linear optics, if the beam propagates in a medium surrounded by the material with the lower index of refraction, propagating beam is reflected from the boundaries of outer media, and when the reflections interfere constructively, the beam forms a guided mode. In nonlinear optics, there is no need in the combination of different media, a special material and an optical beam are sufficient. After the invention of laser in 1960, the phenomenon of nonlinearities became feasible to observe. In 1972, analytic stationary solutions of NLS for Kerr materials were found by Zakharov and Shabat \cite{shabat1972exact} for the case of (1+1)D propagation (i.e. a beam propagating in one direction and diffracting along one dimension), however, for the case of (2+1)D, which are more general, no analytic solution was found and numerically found solitons were shown to be unstable. The main reason for instability was that a soliton is formed at a certain power $P_c$, i.e. for $P\neq P_c$ , it diffracts, or starts to oscillate or even it might undergo catastrophic collapse \cite{snyder1997accessible}. In \cite{chiao1964self}, a stationary solution of (2+1)D NLS was numerically found, nowadays is known as \textit{Townes soliton}, which is illustrated in Figure 2-3.
\begin{figure}[h] %!t
\centering
\includegraphics[width=5in]{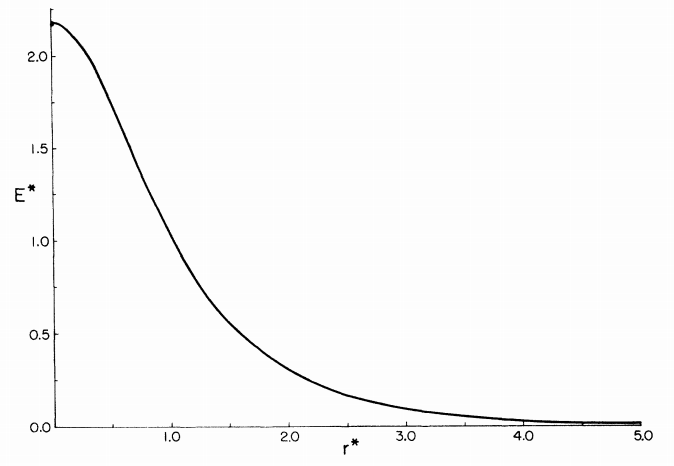}
\caption{Transverse profile of Townes soliton \cite{chiao1964self}}
\label{bench}
\end{figure}
In addition to the instability of spatial solitons in Kerr media, sufficiently high power is needed to create a soliton, it exceeds $1MW/cm^2$ \cite{aitchison1990observation}. As we have mentioned in the previous section, a material with saturable nonlinearity can avoid these problems. In 1974, Ashkin and Bjorkholm \cite{bjorkholm1974cw} provided an evidence of trapping (2+1)D beam in saturable medium. They used cw dye laser as an input beam, and the medium was sodium vapor in vacuum contained inside Pyrex cell. Figure 2-4 shows last 13cm of the cell for two cases: (a) for normal divergence and (b) self-trapping.
\begin{figure}[h] %!t
\centering
\includegraphics[width=5in]{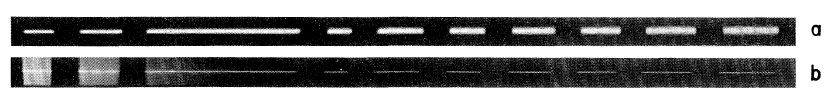}
\caption{Observation of (a) diverging and (b) self-trapped beams \cite{bjorkholm1974cw}}
\label{bench}
\end{figure}
One of their results was that propagation behavior depends on the power of the beam and temperature of the medium. Figure 2-5 shows half power diameter for input beam with power (a) 15 mW, (b) 23 mW and (c) 23 mW with side-arm temperature of 200 degrees of Celsius. There is a significant difference between free-space propagation (solid line) and cases (a), (b) and (c).
\begin{figure}[h] %!t
\centering
\includegraphics[width=4in]{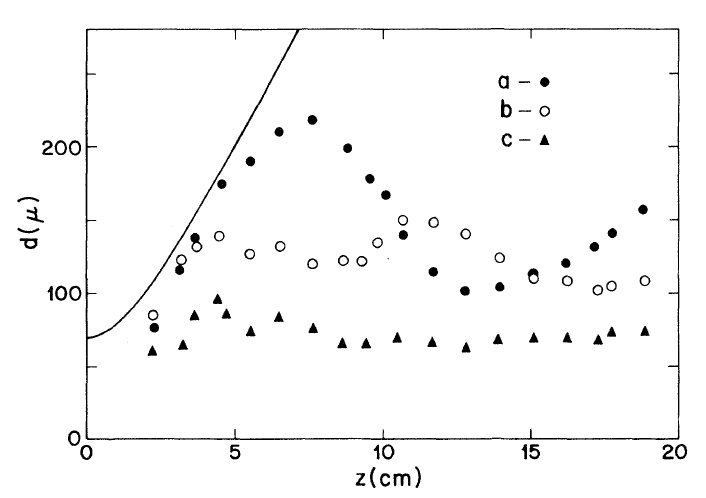}
\caption{Half power diameter of laser beams during propagation in free-space (solid line) and sodium vapor with input power (a) 15 mW, (b) 23 mW, (c) 23 mW with increased temperature of the medium \cite{bjorkholm1974cw}}
\label{bench}
\end{figure}
We observe that as power increases, beam becomes more trapped.
\par
In 1992, a new type of spatial solitons, generated by the photorefractive effect of the medium, were predicted \cite{segev1992spatial}. Efficiency of this effect is that soliton can be formed even at very low powers. A photorefractive soliton is a stationary solution of a wave equation describing a propagation of laser beam in a photorefractive material. 
% This equation is quite hard to explain, however, for an isotropic photorefractive material, a nonlinear material can be modeled through a saturable nonlinearity as in Eq.1.[NEED TO EXPLAIN] 
In 1993, a photorefractive soliton was observed for the first time \cite{duree1993observation}. In all preceding experiments, a very similar apparatus was used: argon-ion laser and a crystal as a photorefractive material. An experiment was performed for two cases: with no voltage in transverse dimension and some applied voltage across the crystal. In \cite{duree1994dimensionality}, an astonishing result was presented, a beam with nonuniform transverse phase can eventually transform to a soliton, whose transverse phase is uniform. To create an input beam with uniform phase, they launched a Gaussian beam through lens so that a waist was at the entrance face.
\begin{figure}[h] %!t
\centering
\includegraphics[width=3in]{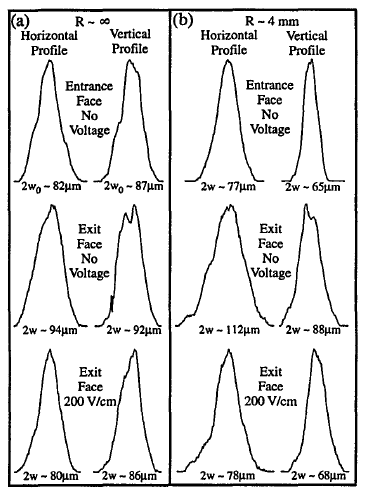}
\caption{Effect of applied voltage on laser beams with (a) uniform and (b) nonuniform input phases \cite{duree1994dimensionality}}
\label{bench}
\end{figure}
Figure 2-6 illustrates above two cases: uniform and nonuniform transverse phases. An external field of 200V/cm was applied. Their observations indicate that the soliton formation takes place at the same value of an external field, regardless of input phase. Figure 2-6(a) is the case when the waist is at the entrance face (i.e. uniform input phase), and diffraction is canceled when external field is applied. Figure 2-6(b) is the case when the radius of curvature is 4mm (i.e. nonuniform input phase), and diffraction is canceled too. In addition, the output phase of the beam with nonuniform input phase is uniform. It might be the manifestation of the stability of photorefractive solitons \cite{segev1994stability}. The above described soliton is one of three types of photorefractive solitons:
\begin{itemize}
\item quasi-steady-state solitons: due to photorefractive effect
\item screening solitons: intensity-dependent screening of an external electric field
\item photovoltaic solitons: no external biasing field
\end{itemize}
A comparison
between these solitons with conventional Kerr-type solitons shows the overwhelming importance of the photorefractive solitons: not
only do optical Kerr solitons require at least 100kW powers\cite{duree1993observation}, but they also do not exist in the bulk(unstable), i.e. they must be launched in a slab waveguide \cite{shih1995observation}. 
\par
There are more experimental evidence presented in 90s. In 1994, a 2D steady-state screening soliton was observed \cite{segev1994steady}. In 1996 \cite{shih1996circular}, an ability of these solitons to waveguide was found. A waveguide was induced by a screening soliton when the soliton beam is on. Number of modes that can be guided by a screening soliton depends on intensity ratio. By producing a $TEM_{01}$ mode, it could be guided by a soliton of intensity ratio of 120. However, when the same mode is launched into a soliton with intensity ratio 3, it is not guided anymore (Figure 2-7).
\begin{figure}[h] %!t
\centering
\includegraphics[width=5in]{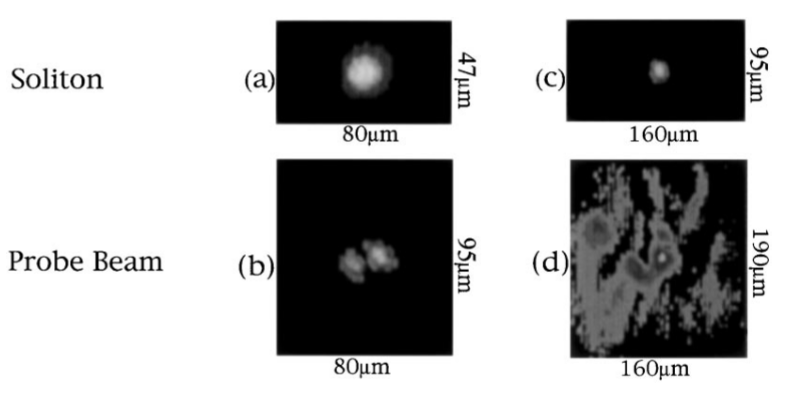}
\caption{Guided TEM$_{01}$ (b) by soliton of intensity ratio 120 (a); unguided TEM$_{01}$ (d) by soliton of intensity ratio 3 (c) \cite{shih1996circular}}
\label{bench}
\end{figure}
\par
The reason for presenting such amount of experimental evidence here, is that during the generation of screening solitons, an induced change of refractive index is analogous to saturable nonlinearity \cite{weilnau2002spatial}. Hence, photorefractive screening solitons are good candidates to check experimentally theoretical conclusions about saturable nonlinearity.
\par
The model describing the propagation of cw laser beam in an isotropic saturable medium is the same as the cubic NLS, except the nonlinear term \cite{yang2002internal}
\begin{equation}
i\frac{\partial\psi}{\partial z}+\Delta_{\perp}\psi-\frac{\psi}{1+|\psi|^2}=0
\end{equation}
To solve it, we use ansatz used in Jianke Yang's \cite{yang2002internal} and Gatz's \cite{gatz1991soliton} papers, namely radially symmetric soliton of the form
\begin{equation}
\psi(r,\theta,z)=u(r)e^{i\omega z}
\end{equation}
Then we get an ODE for $u(r)$
\begin{equation}
u_{rr}+\frac{1}{r}u_r-\omega u-\frac{u}{1+u^2}=0
\end{equation}
with a realistic boundary condition $u(r\rightarrow\infty)\rightarrow 0$. Also, the first derivative of $u$ has to vanish at $r=0$ in order not to have a singular point. To analyze Eq.(2.18), let's consider different asymptotes of $u$. For instance, when $r\gg 1$, Eq.(2.18) can be linearized 
\begin{equation}
u_{rr}+\frac{1}{r}u_r-(1+\omega)u=0
\end{equation}
For $\omega > -1$ and using a specific substitution $x=r\sqrt[2]{1+\omega}$, we get a solution
\begin{equation}
u(r)=\alpha K_0(r\sqrt[2]{1+\omega})
\end{equation}
To numerically find soliton solutions of Eq.(2.18), the shooting method is used \cite{yang2002internal}: fixing $\omega$ and varying $\alpha$ in Eq.(2.20), and by starting from the asymptotic solution and integrating it till $r=0$, $u(r=0,\alpha)$ can be found. 
\begin{figure}[h] %!t
\centering
\includegraphics[width=3in]{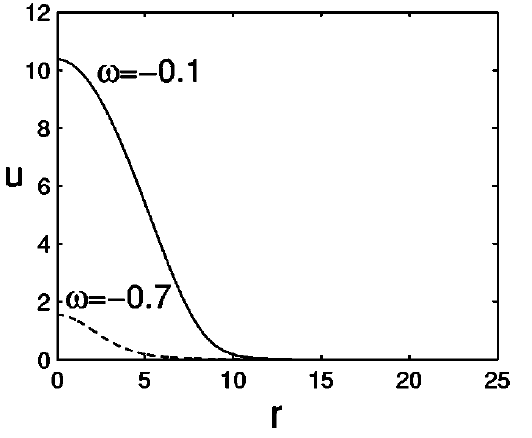}
\caption{Transverse profiles of fundamental solitons with $\omega=-0.1$ (solid) and $\omega=-0.7$ (dashed) \cite{yang2002internal}}
\label{bench}
\end{figure}
If at a certain $\alpha$ $u'(r)$ changes sign, then soliton conditions stated above are satisfied and soliton is found. Using this strategy, it was found by Jianke Yang that for $-1<\omega<0$ there are infinite sequence of soliton solutions. In general, solitons can be categorized according to transverse intensity distributions: fundamental and multi-hump. A fundamental soliton is the one which has an intensity profile consisting from a single peak, and multi-hump solitons consist from a superposition of different modes, which result in several peaks in intensity distribution \cite{ostrovskaya1999stability}. Figure 2-8 illustrates numerically found soliton solutions for $\omega=-0.1$ and $\omega=-0.7$. For $\omega>0$, it was found that $u'(r=0,\alpha)$ does not change its sign for any $\alpha$, hence no fundamental soliton exists there. If $\omega<-1$, for every such $\omega$ a continuous family of solitons were found. However, they are unphysical, because they are described by Bessel functions for large $r$ and have infinite powers. Thus, we restrict our attention to optical solitons with $-1<\omega<0$. 
\par
We know that (2+1)D solitons are unstable in Kerr materials. However, propagation in saturable medium showed to prevent catastrophic collapse. Hence, such materials are useful in analyzing the behavior of solitons. In the next section, stability properties of solitons in saturable medium are presented.

\section{Stability analysis}

A model describing propagation of cw laser beams in (2+1)D Kerr media, or in media with saturable nonlinearity, is nonintegrable and cannot be solved analytically \cite{kivshar2003optical}. In addition to this difficulty, we have to make sure that spatial solitons (or stationary solutions of above models) are stable (or weakly unstable), because only such self-trapped beams can be observed in an experiment \cite{trillo2013spatial}. After the description of optical spatial solitons, we need to explore the stability properties. 
\par
The starting point is our model, a dimensionless generalized NLS equation
\begin{equation}
i\frac{\partial E}{\partial z}+\Delta_{\perp}E+F(I)E=0,
\end{equation}
where $F(I)$ is the nonlinearity related to Kerr-type materials. For instance, for pure Kerr material $F(I)=I$ and for saturable material nonlinearity can be modeled as $\frac{I}{1+I}$. We already know that spatial solitons are found in the generic form as
\begin{equation}
E(x,y,z)=U(x,y)e^{ikz+i\phi(x,y)},
\end{equation}
where a real phase $\phi$ is considered separately, so that $U$ is a real function too. As a result, a closed system of equations for $U$ and $\phi$ is obtained
\begin{align}
\Delta_{\perp}U-kU-(\nabla \phi)^2+F(U^2)U=0 \\
\Delta_{\perp}\phi+2\nabla \phi\nabla \textrm{ln}U=0
\end{align}
To simplify analysis of the above system, we consider different values for $\phi$ separately. For $\phi=0$, it is known that $U$ has to be symmetric in transverse dimensions, i.e. $U=U(r)$ \cite{desyatnikov2005optical}. A particular case $\phi\neq 0$ was investigated by Kruglov and Vlasov \cite{kruglov1985spiral}, where a phase $\phi$ was a multiple of coordinate $\theta=\textrm{tan}^{-1}(y/x)$. By letting $\phi=m\theta$, an equation which has phase dependent stationary solutions can be derived
\begin{equation}
\frac{d^2U}{dr^2}+\frac{1}{r}\frac{dU}{dr}-\frac{m^2}{r^2}U-kU+F(U^2)U=0
\end{equation}
For $r=0$, $U$ has to vanish. From the condition of field univocacy, $m$ is an integer \cite{desyatnikov2005optical}. Such stationary solutions are called \textit{vortex solitons}, and were experimentally found later in \cite{afanasjev1995rotating,skryabin1998dynamics}. 
To derive stability conditions, a linear stability analysis will be used. Assuming that $E_0$ is a stationary solution of Eq.(2.21), its stability can be analyzed through the behavior of a small perturbation $|p|\ll|E_0|$. By substituting a perturbed soliton $E=E_0+p$, Eq.(2.21) can be linearized with respect to $p$
\begin{equation}
i\frac{\partial p}{\partial z}+\Delta_{\perp}p+(F_0+|E_0|^2F_0')p+E_0^2F_0'p^*=0
\end{equation}
where $F_0=F(|E_0|^2)$ and $F_0'=\frac{dF}{dI}|_{I=|E_0|^2}$. Eq.(2.26) describes propagation of initially small perturbation $p$ corresponding to $E_0$. So, if $p$ does not grow with the beam propagation, then $E_0$ is linearly stable. For the case of the soliton in the form $E_0=U(r)e^{im\theta+ikz}$, a corresponding perturbation $p$ should be azimuthally periodic, i.e. it can be represented as a Fourier series
\begin{equation}
p(r,\theta,z)=\sum_{n=-\infty}^{\infty} p_n(r,z)e^{in\theta}
\end{equation}
Substituting it into Eq.(2.26), we obtain an infinite set of systems of equations in the form
\begin{gather}
\left\lbrace i\frac{\partial}{\partial z}+\hat{L}^{+}\right\rbrace p_{m+s}+e^{i2kz}Ap_{m-s}^*=0 \\
\left\lbrace i\frac{\partial}{\partial z}+\hat{L}^{-}\right\rbrace p_{m-s}+e^{i2kz}Ap_{m+s}^*=0
\end{gather}
where $A=U^2F_0'$ and $\hat{L}^{\pm}=\frac{d^2}{dr^2}+\frac{1}{r}\frac{d}{dr}-(m\pm s)^2\frac{1}{r^2}+F_0+A$. Note that $p_{m\pm s}$ are perturbation modes, and they are solutions of the form $p_{m+s}(r,z)=u_s(r)e^{ikz+\gamma_sz}$ and $p_{m-s}(r,z)=v_s^*(r)e^{-ikz+\gamma_s^*z}$, where modes $u_s$ and $v_s$, and complex wavenumber $\gamma_s$ satisfy eigenvalue problem 
\begin{gather}
i\gamma_s
\begin{bmatrix}
u_s \\ v_s
\end{bmatrix}
=
\begin{bmatrix}
k-\hat{L}^+ && -A \\ A && -k+\hat{L}^-
\end{bmatrix}
\begin{bmatrix}
u_s \\ v_s
\end{bmatrix}
\end{gather}
From the expressions of perturbation modes $p_{m\pm s}$, we see that for $\gamma_s$ with positive real part, they grow exponentially, thus, such modes are \textit{instability modes}. Now, an initially small perturbation mode $p_m$ can be represented as $p_m(r,\theta,z)=e^{im\theta+ikz}(u_s(r)e^{is\theta+\gamma_sz}+v_s^*(r)e^{-is\theta+\gamma_s^*z})$. Even though there are many results on stability properties of vortex solitons \cite{desyatnikov2005optical,tikhonenko1995spiraling,chen1997steady,petrov1998observation}, there is no general criterion for stability of vortex solitons such as for fundamental solitons ($m=0$). 
\par 
The primary work on stability of fundamental solitons in saturable medium was in 1978 by Vakhitov and Kolokolov \cite{vakhitov1973stationary}, consequently, \textit{Vakhitov-Kolokolov stability criterion} was derived: a fundamental soliton is linearly stable if its power is an increasing function of soliton propagation constant $k$, i.e. $\frac{dP}{dk}>0$. To be consistent with the work of Jianke Yang in 2002 \cite{yang2002internal}, we use a new parameter $\lambda_s=i\gamma_s$. We also remind ourselves that fundamental solitons exist for $-1<k<0$. Since fundamental solitons are known to be stable in a saturable media, there are internal modes describing periodic oscillations. Quantitatively, it means that $\textrm{Im}(\lambda)=0$. 
% To find all these  modes, we start from the perturbed fundamental soliton
% \begin{equation}
% E(r,\theta,z)=e^{ikz}(\psi(r)+u_s(r)e^{i(\lambda z+s\theta)}+v_s^{*}(r)e^{-i(\lambda^{*}z+s\theta)})
% \end{equation}
Simplicity of stability analysis for fundamental solitons mentioned before was that the square of an operator in Eq-s(2.28-2.29) is self-adjoint when $m=0$, i.e. eigenvalue is purely real or imaginary; however, for vortex solitons it is not true, and eigenvalue may have both nonzero parts. In case of fundamental solitons, $u_s$ and $v_s$ are real. The boundary conditions are
\begin{gather}
u_{0r}(r=0)=v_{0r}=0 \\
u_{s}(r=0)=v_{s}=0, \quad s\neq 0 
% \\ % u_s\rightarrow 0, \quad v_s\rightarrow 0, \quad \textrm{as} \quad r\rightarrow \infty
\end{gather}
Similarly, as in the previous section, for large $r$
\begin{gather}
u_s(r)=K_s(r\sqrt[2]{1+k+\lambda}) \\
v_s(r)=hK_s(r\sqrt[2]{1+k+\lambda})
\end{gather}
% [EXPLAIN THE STRATEGY WITH FIGURES HERE] 
For a particular $s=n$, families of internal modes with varying $k$  were found, which is illustrated in Figure 2-9.
\begin{figure}[h] %!t
\centering
\includegraphics[width=3in]{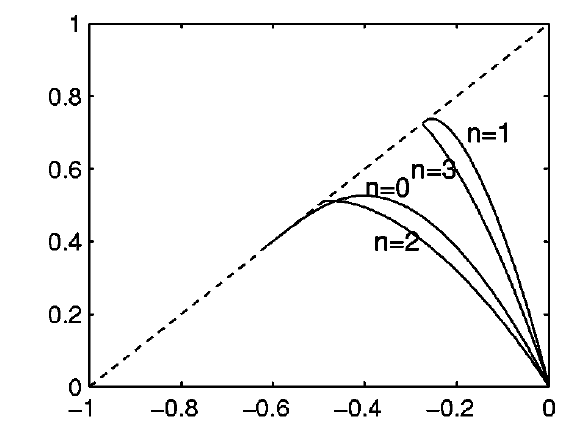}
\caption{Eigenvalues of internal modes $n=0,1,2,3$ vs $k$ \cite{yang2002internal}}
\label{bench}
\end{figure}
Figure 2-9 illustrates that internal modes with $s=0$ and $s=2$ are farthest from continuous spectrum, thus radiation damping of these modes is the slowest, i.e. oscillations resulting from these modes are robust. Also, eigenvalues of the modes approach $0$ when $k$ goes to zero from left. In other words, from the Vakhitov-Kolokolov criterion, internal oscillations of high-power solitons are more robust. Next, the dynamics under the internal modes $n=0$ and $n=2$ is studied, which will be necessary to understand knotting of optical vortices in chapter 3. Firstly, if we consider initial condition
\begin{equation}
E(r,\theta,z=0)=(1+\epsilon)\psi(r,k),
\end{equation}
where $\epsilon$ is some small perturbation parameter, this radially symmetric initial condition will only excite $n=0$ mode (and some radiation), which corresponds to radial stretching of soliton. 
\begin{figure}[h] %!t
\centering
\includegraphics[width=5in]{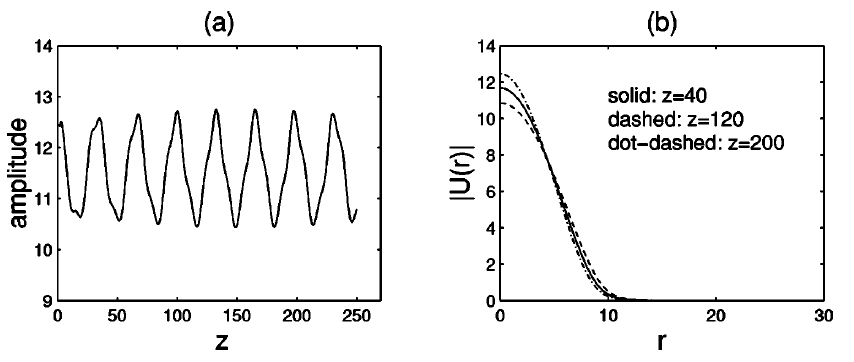}
\caption{(a) Evolution of beam center's amplitude; (b) Amplitude profiles at 3 different distances \cite{yang2002internal}}
\label{bench}
\end{figure}
Figure 2-10(a) shows the amplitude at the center during propagation. We see that it persists a very robust oscillation. Figure 2-10(b) shows amplitude profiles for different propagation distances. 
\begin{figure}[h] %!t
\centering
\includegraphics[width=4in]{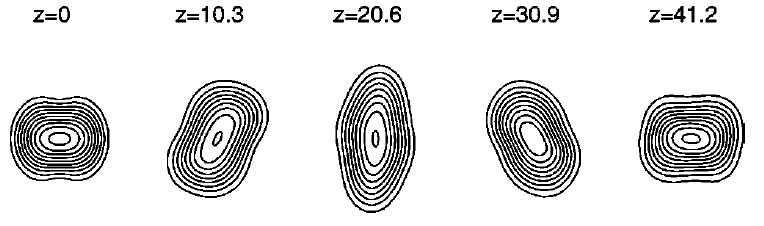}
\caption{Effect of $n=2$ perturbation mode on fundamental soliton with $k=-0.1$ at different $z$ \cite{yang2002internal}}
\label{bench}
\end{figure}
It shows that radiation emission is very small, thus, it is expected that it will oscillate for a very long distance. For the case of $n=2$ internal mode, the initial condition is
\begin{equation}
E(r,\theta,z=0)=\psi(r,k)+\epsilon(u_2(r)e^{i2\theta}+v_2(r)e^{-i2\theta})
\end{equation}
Figure 2-11 illustrates the contours of $|E|$ at five different distances. It shows that a fundamental soliton is stretched along some direction as it propagates, which looks like it rotates. $n=2$ mode corresponds to a twist of soliton. We see that, above simulations have similar $breathing$ behavior.
\par
In this chapter, we presented an equation describing propagation of slowly varying envelope of optical field in Kerr-type materials. Then, we discussed about spatial optical solitons, which are stationary solutions of NLS equation, together with their stability properties in saturable medium. In particular, we are interested specifically in two modes, which will  help us to construct vortex knots around perturbed fundamental soliton propagating in saturable medium. Such system will be very similar to the one, where knotting of vortex lines seems spontaneous, and will be described in chapter 3.
%% This is an example first chapter.  You should put chapter/appendix that you
%% write into a separate file, and add a line \include{yourfilename} to
%% main.tex, where `yourfilename.tex' is the name of the chapter/appendix file.
%% You can process specific files by typing their names in at the 
%% \files=
%% prompt when you run the file main.tex through LaTeX.
\chapter{Vortex lines topology}
In this chapter we describe  topology of optical vortex lines in linear and nonlinear media. As a main part, we focus on a problem of identifying vortex knots around a perturbed fundamental soliton in nonlinear saturable medium.
\section{Optical vortex lines in linear media}
Development of topology, as a science, started at the beginning of 20th century \cite{croom2016principles}. Initially, it was just a branch of mathematics. However, as applied sciences progressed, topology became an interdisciplinary science. A research of DNA structures shows that knotting of DNA molecules is directly related to replication and recombination processes \cite{bates2005dna}. Another example is a role of topology in polymer science, knotting of polymers is proven to be necessary for crystallization properties \cite{de1979scaling}. 
% Looking at previous examples, it is understandable why research in this field is still going on. 
In \cite{water_knot}, a creation of trefoil vortex knot propagating in water was reported. There are several scientific works done in the field of optics as well, where scientists were able to embed knotted vortex lines in a laser beam \cite{knot_link_mono,isol,dark_threads,vortex_knots}. First example is LG high-order modes, where a vortex line on the optical axis is located. In \cite{knot_link_mono}, a creation of knotted vortex lines in a perturbed superposition of $m$-Bessel beams of the form
\begin{equation}
F_{mb}(R,z)=\exp(im\phi)J_m(bR)\exp(iz\sqrt[]{1-b^2})
\end{equation}
\begin{figure}[h] %!t
\centering
\subfloat[]{{\includegraphics[width=2.4in]{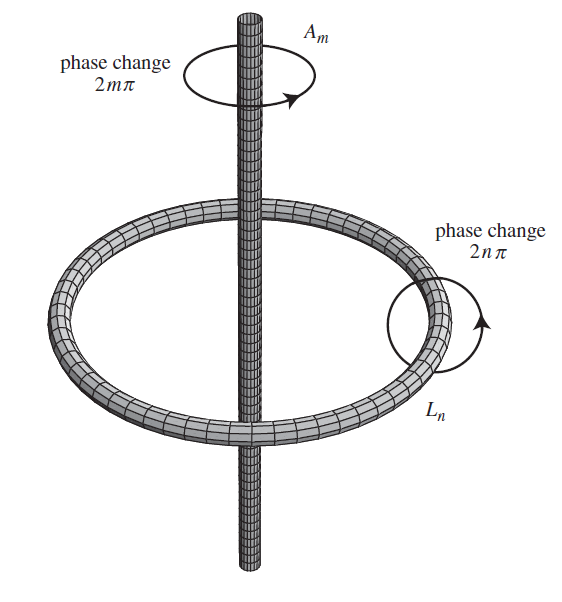} }}%
\qquad
\subfloat[]{{\includegraphics[width=2in]{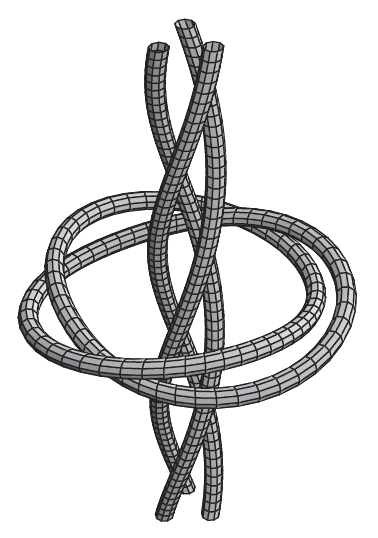} }}%
\caption{(a) Vortex loop of charge $n$ and a vortex line of charge $m$. (b) Perturbed structure in (a) for $n=2$ and $m=3$, resulting in vortex trefoil knot threaded by vortex 3-stranded helix \cite{knot_link_mono}}
\label{bench}
\end{figure}
was considered. Such beams satisfy Helmholtz equation, and possess an optical vortex of charge $m$ on $z$-axis. A suitable superposition of such beams was chosen to create a vortex loop of charge $n$, illustrated in Figure 3-1(a). In particular, a beam with a vortex loop of charge 2 and axial optical vortex line with charge 3 was considered. As a result of perturbations with circular symmetry, a (2,3) torus knot (trefoil knot) threaded with 3-stranded helix was constructed. In general, if $m$ and $n$ are co-prime, then a ($m$,$n$) torus knot threaded by $m$-stranded helix is formed. If ($m$,$n$)=$p$($m_0$,$n_0$), where ($m_0$,$n_0$)=1, then $p$ linked ($m_0$,$n_0$) knots threaded by $m$-stranded helix are formed \cite{knot_link_mono}. In \cite{isol}, a different method of vortex knot construction was presented. Initially, a complex polynomial whose roots form helical and pigtail braids were constructed. A polynomial describing a helical braid with $n$ number of repeats is
\begin{equation}
q_{helix}=u^2-v^n,
\end{equation}
where $v=e^{ih}$, $h$ is the $2\pi$ periodic height parameter. Then, using a specific mapping of $u$ and $v$ into real 3D space
\begin{equation}
u(\boldsymbol r)=[r^2+z^2-1+2iz]/(r^2+z^2+1), \quad v(\boldsymbol r)=2re^{i\phi}/(r^2+z^2+1),
\end{equation}
a set of knotted vortex lines, determined by the braid, can be obtained. 
\begin{figure}[h] %!t
\centering
\subfloat[]{{\includegraphics[width=2in]{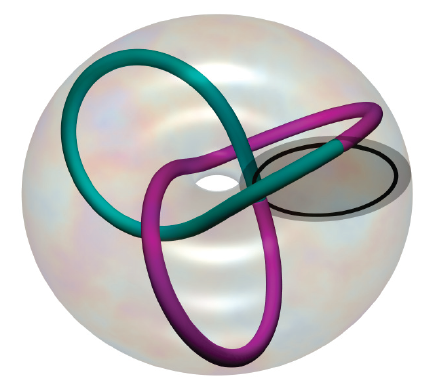} }}%
\qquad
\subfloat[]{{\includegraphics[width=2.5in]{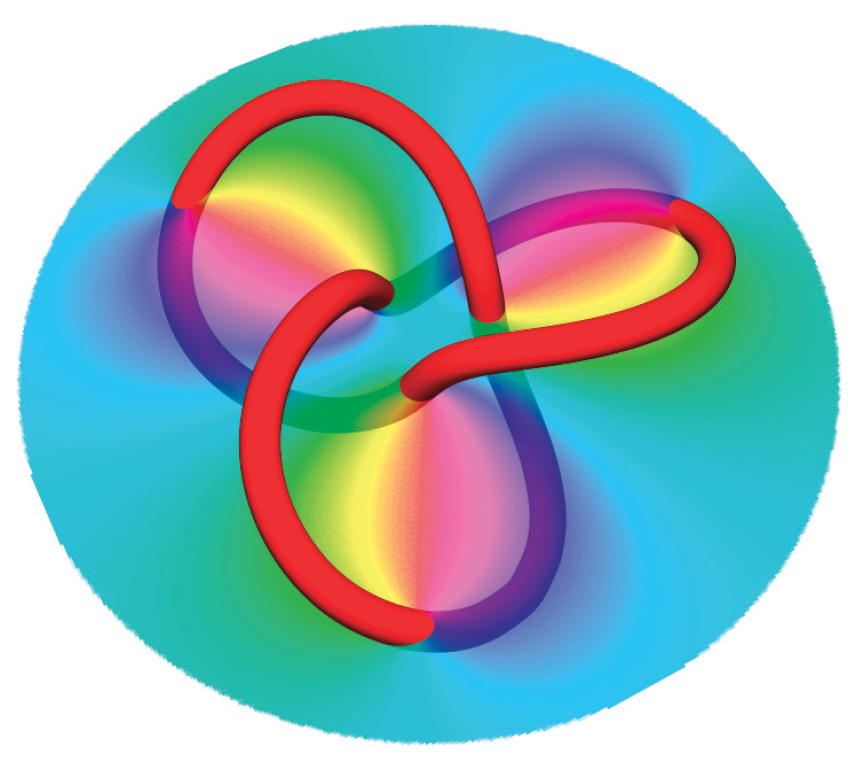} }}%
\caption{(a) Mapping helical braid by $u(\boldsymbol r),v(\boldsymbol r)$, resulting in knotted nodal set; (b) Vortex trefoil knot in optical field of the form Eq.(3.5), multiplied by appropriate Gaussian \cite{isol}}
\label{bench}
\end{figure}
A $q_{helix}$ has a set of roots, which forms a (2,$n$) torus knot. Figure 3-2(b) illustrates a braid transformed into (2,3) torus knot. A numerator of $q$ after applied transformation $u(\boldsymbol r)$ and $v(\boldsymbol r)$ is a polynomial
\begin{equation}
(r^4-1)(r^2-1)-8r^3e^{3i\phi}+4iz(r^4-1)+z^2(3r^4-6r^2-5)+8iz^3r^2+z^4(3r^2-5)+4iz^5+z^6
\end{equation}
Corresponding polynomial which coincides with Eq.(3.4) at $z=0$ and satisfies paraxial equation is
\begin{equation}
(r^4-1)(r^2-1)-8r^3e^{3i\phi}+2iz/k(9r^4-4r^2-1)-8(z/k)^2(9r^2-1)-48i(z/k)^3
\end{equation}
Even though such polynomial solutions were found, they diverge as $x,y\rightarrow \infty$. To avoid it, corresponding polynomial at $z=0$ is multiplied with a Gaussian of width $w$, so that the same vortex knot remains for sufficiently large $w$. To realize it in experiment, a spatial light modulator was used, which can imprint the desired phase distribution into the Gaussian beam. Since techniques to manipulate LG modes experimentally are well-developed, a desired optical field is decomposed into LG modes. However, there is a problem of detection of vortex lines, for instance, to distinguish optical vortex from the low intensity region. To overcome this problem, optimization algorithm was applied. The main idea of this algorithm is to vary coefficients of LG modes, so that contrast was increased and the resultant topology of vortex lines remained the same as before (Figure 3-3).
\begin{figure}[h] %!t
\centering
\includegraphics[width=4in]{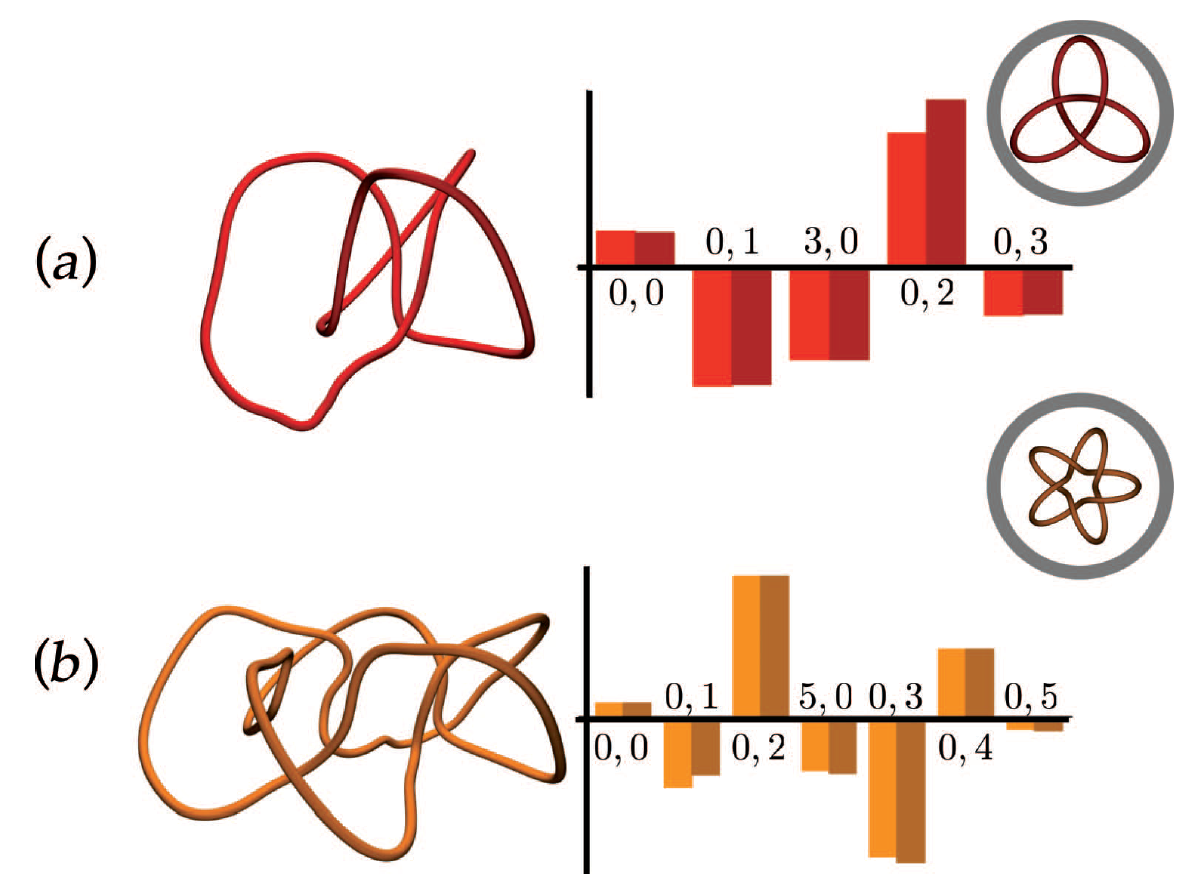} 
\caption{Optimized (darker) and unoptimized (lighter) coefficients of LG modes, which decompose optical field containing (a) trefoil and (b) cinquefoil vortex knots \cite{knot_tangle}}
\label{bench}
\end{figure}

In \cite{knot_tangle}, in addition to possible embedding of vortex knots in a laser beam, vortex lines in optical speckle were considered. Physically, \textit{optical speckle} is produced as an interference of reflections from some rough surface of optical field , therefore, can be considered as a random field. It was shown in \cite{speckle}, that random superposition of many plane waves behaves similar to optical speckle. In \cite{fractal} was shown that in large size volumes, around 25\% of total length of vortex lines are closed loops (Figure 3-4).
\begin{figure}[h] %!t
\centering
\includegraphics[width=4in]{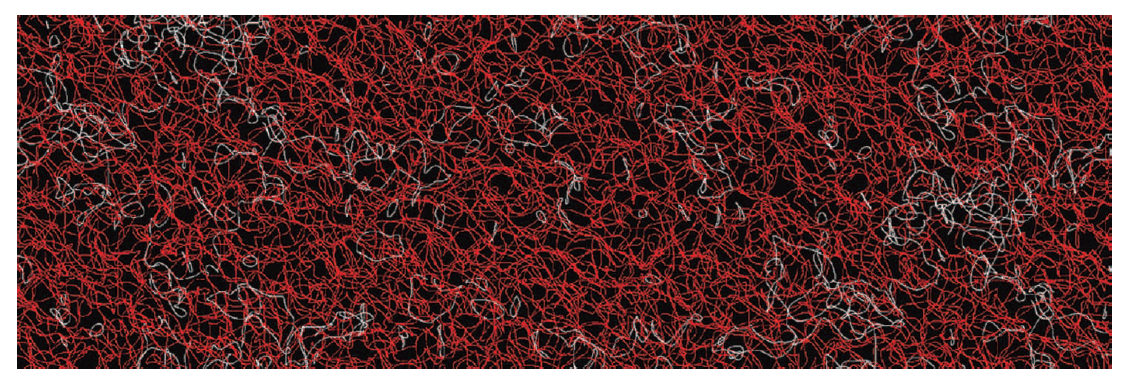} 
\caption{Random optical field containing vortex loops (white) and infinite vortex lines (red) within a particular volume \cite{knot_tangle}}
\label{bench}
\end{figure}
\par
Above we presented examples of knotted optical vortex lines in free space, i.e. linear medium. In the next section, we present similar phenomenon observed in nonlinear medium, which will lead to our main problem of identifying vortex knots around a soliton in nonlinear saturable medium.
\section{Optical vortex lines in nonlinear media}
In this section, optical fields in nonlinear optical media and topology of corresponding vortex lines will be considered. The difference between these media is the behavior of refractive index; refractive index of nonlinear media depends on intensity of optical field, which might result in self-focusing effect. Figure 3-5 illustrates a Gaussian beam propagating in linear (a) and saturable nonlinear media (b).
\begin{figure}[h] %!t
\centering
\includegraphics[width=4.5in]{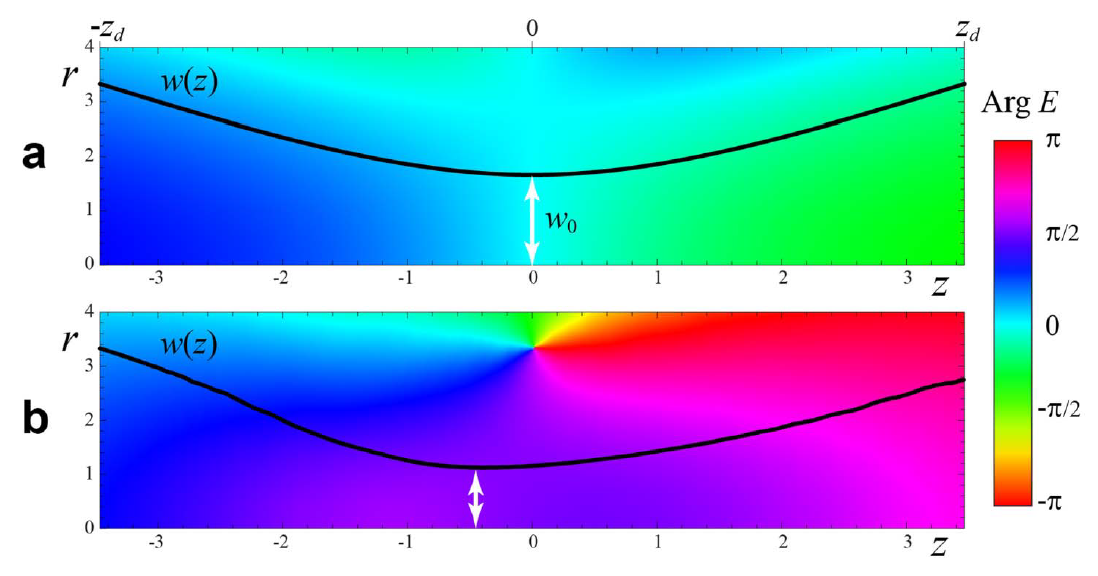} 
\caption{A Gaussian beam in linear (a) and nonlinear (b) media \cite{desyatnikov2012spontaneous}}
\label{bench}
\end{figure}
In Figure 3-5(a), $w_0$ is a value of waist and black line indicates half of the center's intensity at particular $z$ (HWHM). A main difference between (a) and (b), as we might expect, is self-focusing effect, which resulted in the shift of waist location and its decreased value. There appeared a point in longitude plane, where all the phase colors meet. This is an optical vortex. Because of symmetry, there is a vortex loop in transverse plane. 
\par
% A difference from linear media is that knotting of vortex lines can be created without appropriate superposition of waves, in contrast with examples stated in previous section.
Fundamental solitons in saturable medium also have Gaussian like amplitude profile \cite{yang2002internal}. In contrast with Figure 3-5(b), they do not possess optical vortices at all. 
% Also, we know that in \cite{yang2002internal}, as a result of internal modes a breathing effect was observed. However, there is an additional topological effect which was not mentioned there (will be presented soon). 
To create optical vortices around fundamental soliton, a perturbation can be introduced. 
\par
Assume that $E(x,y,z)=A(r)e^{ikz}$ is a stationary solution of dimensionless NLS equation, then, as in \cite{desyatnikov2012spontaneous}, a perturbation of the form
\begin{equation}
E(x,y,z=0)=A(\sqrt[]{x^2/a^2+y^2/b^2})e^{i\Theta xy}/\sqrt[]{ab}
\end{equation}
can be considered, which preserves soliton power; $\Theta$ parameter corresponds to phase twist \cite{suppression_2010,suppression_2011} and $a,b$ correspond to stretching.
\begin{figure}[!htpb] %!t
\centering
\includegraphics[width=4.5in]{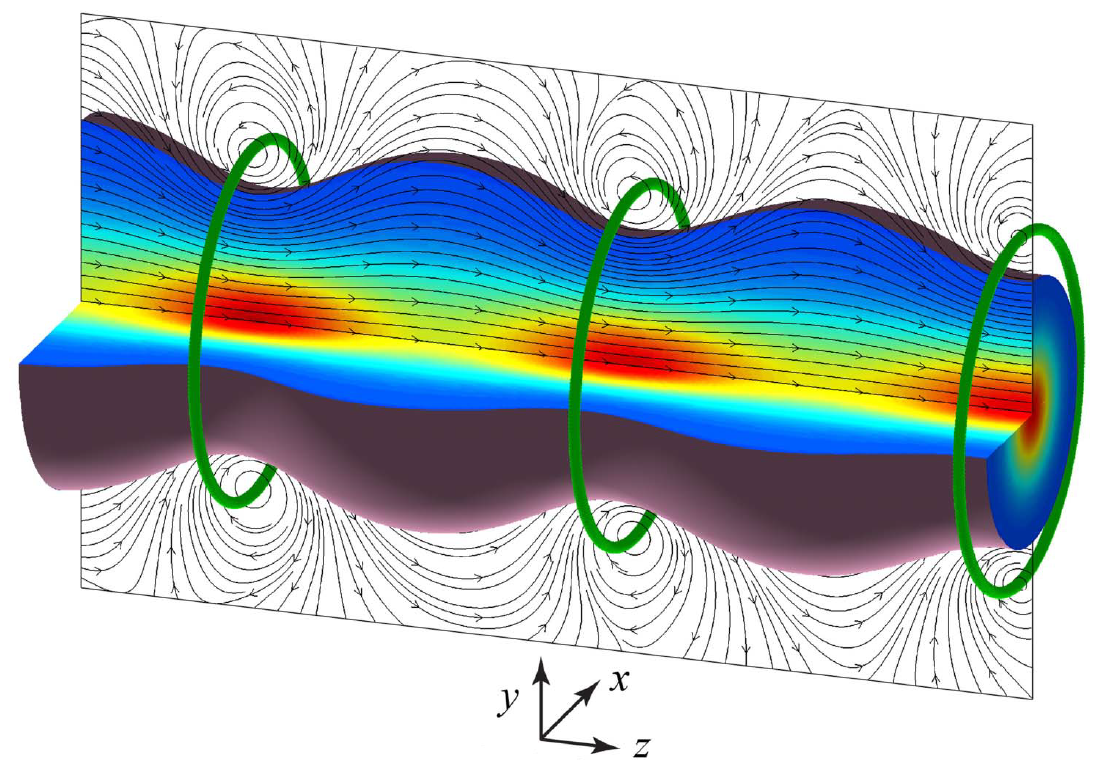} 
\caption{Perturbed soliton with $a=b=1.1$ and $\Theta=0$ \cite{desyatnikov2012spontaneous}}
\label{bench}
\end{figure}
\begin{figure}[!htpb] %!t
\centering
\includegraphics[width=4.5in]{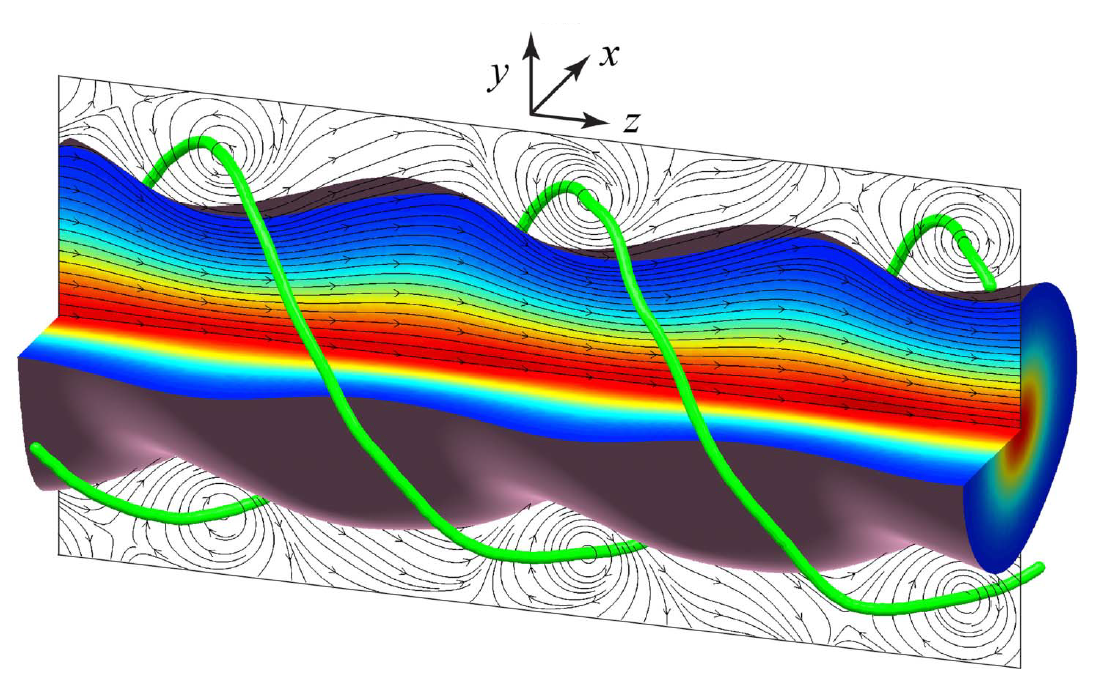} 
\caption{Perturbed soliton with $a=1.42$, $b=0.87$ and $\Theta=0.05$ \cite{desyatnikov2012spontaneous}}
\label{bench}
\end{figure}
\begin{figure}[!htpb] %!t
\centering
\includegraphics[width=4.3in]{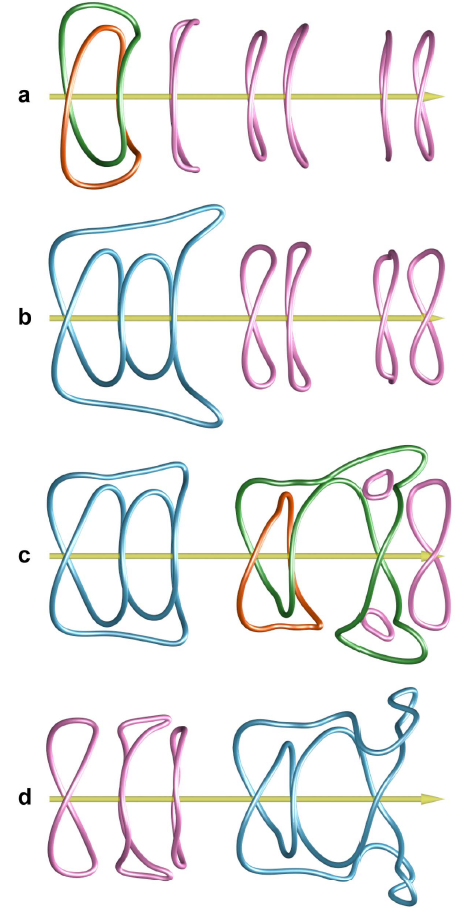} 
\caption{Vortex lines around perturbed soliton with $k=-0.2$, $a=1.1$, $b=1.21$ and different $\Theta$: (a) 0.005; (b) 0.01; (c) 0.013; (d) 0.015 \cite{desyatnikov2012spontaneous}}
\label{bench}
\end{figure}
\newpage
 Figure 3-6 illustrates soliton stretching, i.e. $a=b=1.1$ and $\Theta =0$. When $a\neq b$ and $\Theta$ is nonzero, a different scenario for knotting occurs, as illustrated in Figure 3-7. Producing such perturbed beam is experimentally feasible, as stated in \cite{courtial1997gaussian}. In \cite{desyatnikov2012spontaneous}, set of numerical simulations for certain values of power $P$, $a$, $b$ and various $\Theta$ is presented (Figure 3-8). 
\par
We note that Figures 3-6 and 3-7 are very similar to Figure 2-11, which resulted from effects of perturbation modes. The possible way to understand a mechanism responsible for knotting of vortex lines is provided in the next section.
\section{Knotted vortex lines around perturbed fundamental soliton}
In numerical simulations presented in the previous section, topology was shown to be robust with respect to small changes in values of $\Theta$ \cite{desyatnikov2012spontaneous}. It might give us a clue about which perturbation modes are excited during stretching and twisting. From Figure 2-9, we know that oscillations resulting from modes $s=0$ and $s=2$ are the most robust \cite{yang2002internal}. Hence, it is reasonable to consider a superposition of fundamental soliton with these modes and compare with results from previous section.
\par
As was discussed in section 2.3, a soliton perturbed by internal mode is of the form
\begin{equation}
E(x,y,z)=e^{ikz}\left\lbrace A(r)+u_s(r)[\epsilon_1e^{is\phi}+\epsilon_2e^{-is\phi}]e^{i\omega_sz}+v_s(r)[\epsilon_1e^{-is\phi}+\epsilon_2e^{is\phi}]e^{-i\omega_sz}\right\rbrace
\end{equation}
$u_s$ and $v_s$ are envelopes of perturbation modes. By normalizing perturbation modes, so that $max(|u_s|,|v_s|)=max(A)$, magnitudes of $\epsilon_{1,2}$ can be treated as strengths of perturbations. Envelopes of perturbation can be found numerically, from the boundary conditions presented in Section 2.3. 
\begin{figure}[!htpb] %!t
\centering
\includegraphics[width=4.3in]{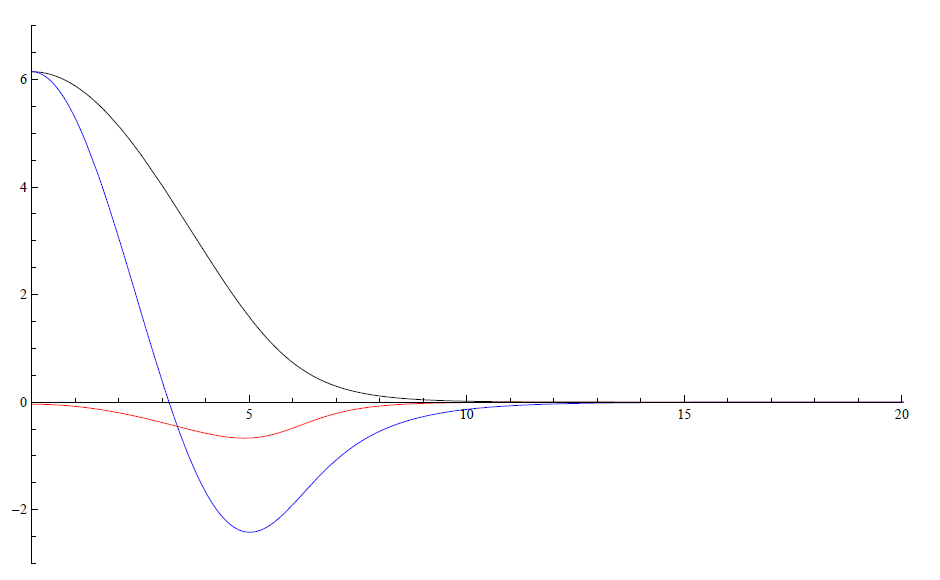} 
\caption{Profiles of $A(r)$ (black), $u_0(r)$ (red), $v_0(r)$ (blue) for $k=-0.2$, $A_0=6.14465$, $\omega_0=0.382777$ and $h=0.218418$}
\label{bench}
\end{figure}
\begin{figure}[!htpb] %!t
\centering
\includegraphics[width=4.3in]{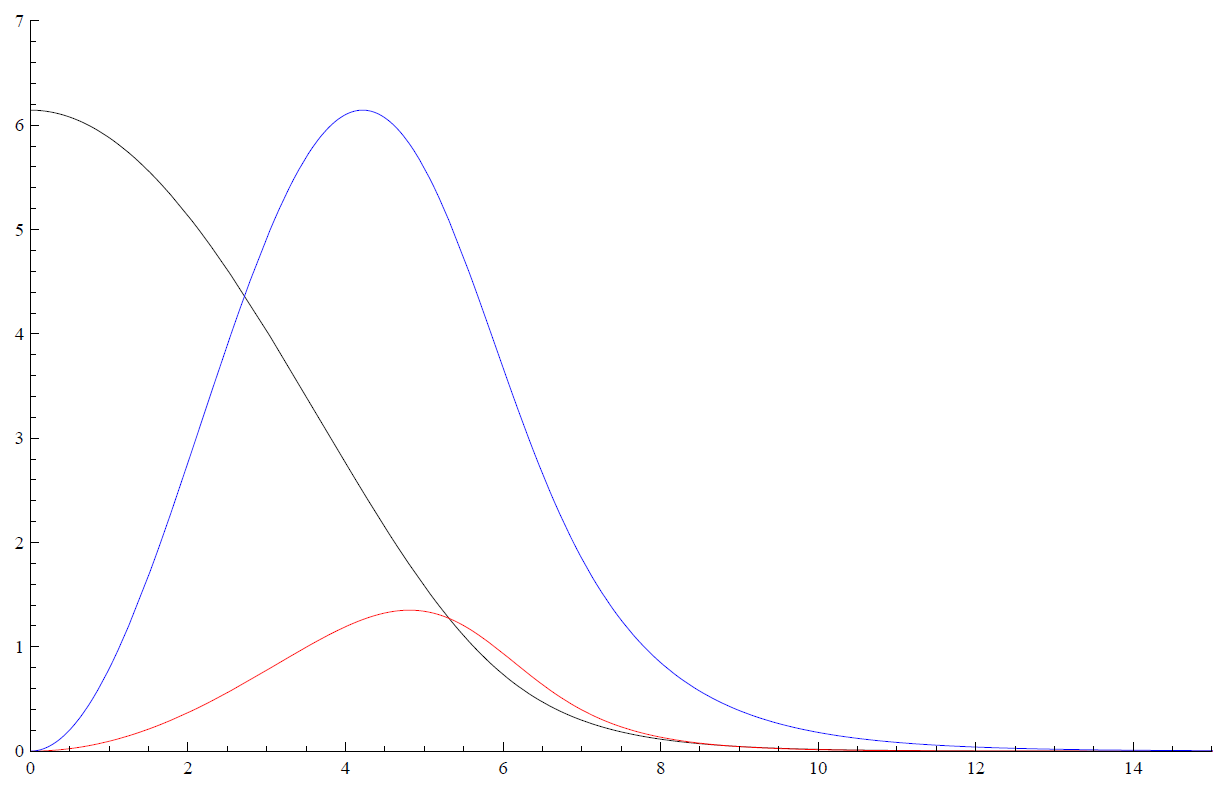} 
\caption{Profiles of $A(r)$ (black), $u_2(r)$ (red), $v_2(r)$ (blue) for $k=-0.2$, $A_0=6.14465$, $\omega_2=0.31843$ and $h=0.31689$}
\label{bench}
\end{figure}
% Since internal modes are eigenmodes, an appropriate superposition may construct any possible perturbation, i.e. the perturbation introduced in Section 3.2 can be expanded in terms of internal modes too.
\par
Firstly, we consider effect of these modes separately. $C_1(r,z)$ represents a superposition of fundamental soliton and perturbation mode $s=0$. Optical vortices are set of points $\lbrace r_*,z_*\rbrace$, which satisfy $C_1(r_*,z_*)=0$.
\begin{equation}
C_1(r_*,z_*)=A(r_*)+(\epsilon_1+\epsilon_2)\left[u_0(r_*)e^{i\omega_0z_*}+v_0(r_*)e^{-i\omega_0z_*}\right]
\end{equation}
Let $\epsilon_1+\epsilon_2=\varepsilon e^{i\delta}$, then $C_1(r_*,z_*)=0$ can be transformed into a 4th order polynomial of $x=\textrm{cos}(\omega_0z_*)$
\begin{equation}
\begin{split}
[16u_0^2v_0^2] x^4+[16Qu_0v_0(u_0+v_0)\textrm{cos}(\delta)] x^3 + \\
[4Q^2(u_0+v_0)^2\textrm{cos}^2(\delta)+8Q^2u_0v_0+8u_0v_0(u_0-v_0)^2+4Q^2(u_0-v_0)^2\textrm{sin}^2(\delta)] x^2 + \\
[4Q^3(u_0+v_0)\textrm{cos}(\delta)+4Q(u_0+v_0)(u_0-v_0)^2\textrm{cos}(\delta)] x + \\
[Q^2+(u_0-v_0)^2]^2-4Q^2(u_0-v_0)^2\textrm{sin}^2(\delta)=0,
\end{split}
\end{equation}
where $Q=A/\varepsilon$. The case $\varepsilon=0.1$ and $\delta=\pi/4$ was already considered in \cite{desyatnikov2012spontaneous}, and it was in good agreement with Figure 3-8(a). An algorithm solving this polynomial is already written, and we plan to work on it later, by varying strength $\varepsilon$, and analyzing corresponding optical vortices. 
\par
A superposition of soliton with $s=2$ mode, will result in creation of vortex spirals \cite{desyatnikov2012spontaneous}. We will work on deriving a similar polynomial, however, because of an additional parameter $\phi$, solving this polynomial will be a more difficult task. When $s=0$ and $s=2$ modes are considered simultaneously, optical vortices are the roots of quadratic equation for $t=e^{i2\phi}$ 
\begin{equation}
C_2(r_*,z_*)t^2+C_1(r_*,z_*)t+C_0(r_*,z_*)=0, 
\end{equation}
where $C_{2,0}=\epsilon_{1,2}u_2(r_*)e^{i\omega_2z_*}+\epsilon_{2,1}v_2(r_*)e^{-i\omega_2z_*}$. 
% Figure 3-11 illustrates numerical simulations for various values of $\epsilon_{1,2}$.
% \begin{figure}[!htpb] %!t
% \centering
% \includegraphics[width=4.6in]{3-11.png} 
% \caption{Vortex lines around fundamental soliton with $k=-0.2$, superimposed with perturbation modes $n=0,2$. Values of parameters: $\epsilon_0=0.1$ and $\epsilon_{1,2}=\alpha_{1,2}e^{i\pi/4}$, where (a) $\alpha_1$=0.1, $\alpha_2$=0; (b) $\alpha_1$=0.15, $\alpha_2$=0; (c) $\alpha_1$=0.15, $\alpha_2$=0.015; (d) $\alpha_1$=0.15, $\alpha_2$=0.045.  Length of a yellow arrow represents one period $2\pi/(\omega_0-\omega_2)\approx 97$.}
% \label{bench}
% \end{figure}
% We observe that Figure 3-11(a-b) are very similar to Figures 3-6 and 3-7. In Figure 3-11(c-d), interplay of stretching and elliptic twist resulted in more complex structures, such as knots and links.
This is not an ordinary polynomial for two reasons: coefficients $C_{0,1,2}$ are not constants and desired roots have to be of unity magnitude.
% Initially, we define a family of polynomials $\lbrace f(r_*,z_*,\phi)\rbrace$, where $\lbrace r_*,z_*\rbrace$ are fixed for particular member of this family. Hence, any polynomial of this family has constant coefficients. 
At particular $\lbrace r_*,z_*\rbrace$, there might be 0, 1 or 2 roots $t$ with magnitude unity. 
\par
When there are 2 roots of magnitude unity, say $t_{1,2}=e^{i2\phi_{1,2}}$, using Vieta's theorem, it is concluded that $|C_2|=|C_0|$ and $|C_1|/|C_0|=2|\textrm{cos}(\phi_1-\phi_2)|$. These conditions will provide us with a constraint on $\epsilon_{1,2}$. 
\par
The case, when there is one root of unity magnitude, was considered in \cite{desyatnikov2012spontaneous}. By varying $\epsilon_{1,2}$, they were able to construct every knot from Figure 3-8.
\par
Analyzing the case, when there is no root of unity magnitude, will provide us with information for which $\epsilon_{1,2}$, optical vortices will not occur. 
\par
We have presented a physical system, where particular phenomenon cannot be directly predicted. To understand the mechanism underlying this phenomenon, a similar system was constructed, where we tackled with a mathematical problem. To our knowledge, a full analysis of this problem is still not presented. We considered simplified versions of this problem, and already have a plan to attack it in the future. We will be working on it later, with a more detailed plan presented in the last chapter. Solving this problem, might be a step forward in analyzing another physical system, such as quantum turbulence in superfluids.

% Since $\epsilon_{1,2}$ are free parameters, for fixed $\lbrace r_*,z_*\rbrace$, previous condition will not hold in general. Hence, there is only one root $t$ for particular $\lbrace r_*,z_*\rbrace$. (LIAR)

\newpage

%% This is an example first chapter.  You should put chapter/appendix that you
%% write into a separate file, and add a line \include{yourfilename} to
%% main.tex, where `yourfilename.tex' is the name of the chapter/appendix file.
%% You can process specific files by typing their names in at the 
%% \files=
%% prompt when you run the file main.tex through LaTeX.
\chapter{Summary and outlook}
In this work, we combined some knowledge from different disciplines, such as knot theory, laser physics and singular optics, in order to present the physical system of our interest. Our system is an optical soliton in nonlinear saturable medium. When optical soliton is elliptically stretched and has a twisted phase, spontaneous knotting of optical vortices around the soliton is observed. It is possible to construct a similar system, using a superposition of optical soliton with perturbation modes. $s=0,2$ perturbation modes were sufficient to reconstruct similar structure of optical vortices around the soliton.  Our first step was to analyze a simpler problem, where effect of $s=0$ perturbation mode was only considered, and it led us to 4th order polynomial, which will be addressed in more details later. The case, where both modes are considered, leads to a mathematical problem, for which we are still looking for the ways to solve. To our knowledge, there is no literature where such problem was solved.
\par
Since there is no general theory of solving such problems, we plan to tackle it numerically. Firstly, we will try to identify a domain of strengths of perturbation modes, for which knotting of optical vortices takes place. Then, we plan to create an algorithm, which will divide these domains into subdomains, in each of which the same knotting behavior is presented. At this stage, we also need to apply an algorithm which distinguishes knots. As our final step, we will find and explain a correspondence between above two systems, i.e. which values of strengths of perturbation modes correspond to specific simultaneous action of elliptic stretching and phase twist.
\par
Solving our problem promises to explain similar occurrence of knotted optical vortices around vortex solitons \cite{biloshytskyi2017solitons}. For a vortex soliton with charge $m=1$, there was observed two vortex rings per period, which gives a hint for correspondence between topological charge and number of vortex rings. 
\begin{figure}[!htpb] %!t
\centering
\includegraphics[width=5in]{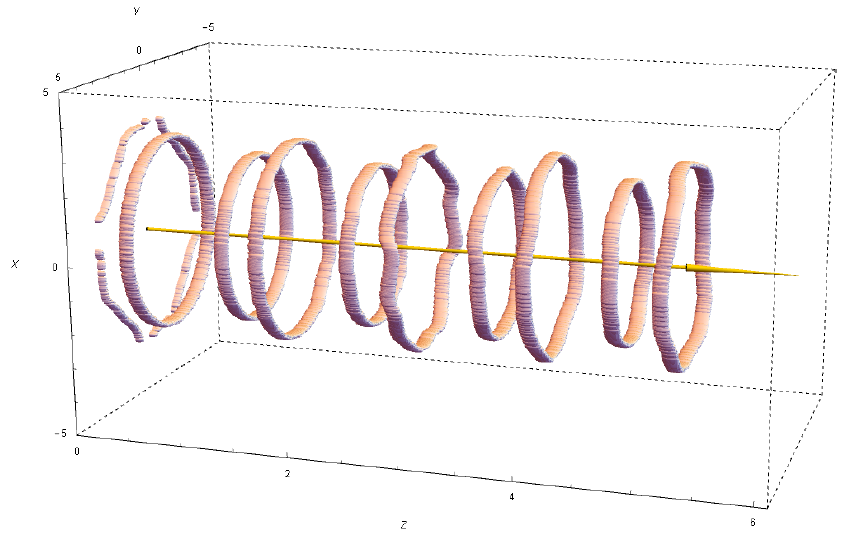} 
\caption{Observation of two vortex rings per period around vortex soliton of charge 1 [by Volodymyr Biloshytskyi]}
\label{bench}
\end{figure}
We plan to check the hypothesis, that if a vortex soliton of charge $m$ is perturbed, then $m+1$ vortex rings per period occur. Later, we will be motivated to expand our system. What is the structure of optical vortices, if perturbed soliton excites higher order modes? Answer to this question might be a step forward in understanding of quantum turbulence in superfluids, where complex dynamics of vortex filaments is observed \cite{barenghi2001quantized}. The last, but not the least, numerical solution of our mathematical problem might give other scientists a clue how to tackle it analytically, which will give a rise to new mathematics.

\appendix %put appendix if you want
%% This defines the bibliography file (main.bib) and the bibliography style.
%% If you want to create a bibliography file by hand, change the contents of
%% this file to a `thebibliography' environment.  For more information 
%% see section 4.3 of the LaTeX manual.

% \begin{singlespace}
% \bibliography{main}
% \bibliographystyle{phys}
% \end{singlespace}

\printbibliography[title={References}]
\end{document}